\documentclass[aps,superscriptaddress,nopacs,nofootinbib,eqsecnum,prd,notitlepage,twocolumn]{revtex4-1} 

\usepackage{graphics} 
\usepackage{epsf}  
\usepackage{eucal} 
\usepackage{amssymb}
\usepackage{amsmath} 
\usepackage{graphicx} 
\usepackage{epstopdf} 
\usepackage{bm}
\usepackage{dcolumn} 
\usepackage{epsfig} 
\usepackage{hyperref}
\usepackage{flafter} 
\usepackage[usenames,dvipsnames]{color}

\newcommand{\be}{\begin{equation}}
\newcommand{\ee}{\end{equation}}
\newcommand{\bea}{\begin{eqnarray}}
\newcommand{\eea}{\end{eqnarray}}

\begin{document}

\title{Inflaton dark matter from incomplete decay}

\author{Mar Bastero-Gil} \email{mbg@ugr.es} \affiliation{Departamento
  de F\'{\i}sica Te\'orica y del Cosmos, Universidad de Granada,
  Granada-18071, Spain}

\author{Rafael Cerezo} \email{cerezo@ugr.es} \affiliation{Departamento
  de F\'{\i}sica Te\'orica y del Cosmos, Universidad de Granada,
  Granada-18071, Spain}

\author{Jo\~ao G. Rosa}
\email{joao.rosa@ua.pt} 
\affiliation{Departamento de F\'{\i}sica da Universidade de Aveiro and CIDMA, Campus de Santiago, 3810-183 Aveiro, Portugal} 

\begin{abstract}
We show that the decay of the inflaton field may be incomplete, while nevertheless successfully reheating the universe and leaving a stable remnant that accounts for the present dark matter abundance. We note, in particular, that since the mass of the inflaton decay products is field-dependent, one can construct models, endowed with an appropriate discrete symmetry, where inflaton decay is kinematically forbidden at late times and only occurs during the initial stages of field oscillations after inflation. We show that this is sufficient to ensure the transition to a radiation-dominated era and that inflaton particles typically thermalize in the process. They eventually decouple and freeze out, yielding a thermal dark matter relic. We discuss possible implementations of this generic mechanism within consistent cosmological and particle physics scenarios, for both single-field and hybrid inflation.

\end{abstract}

\date{\today}

\maketitle


\section{Introduction}

Inflaton and dark matter candidates in particle physics models share several common features, both being typically assumed to be weakly interacting
and neutral fields. The inflaton scalar field  must interact weakly with itself and other degrees of freedom in order to ensure the required flatness
of the associated scalar potential, which could be spoiled by large radiative corrections \cite{inflation}. Similarly, dark matter particles should
form a stable non-relativistic and non-luminous fluid at late times that accounts for the observed galaxy rotation curves \cite{rotationcurves} and
the large scale structure in the universe as inferred from Cosmic Microwave Background \cite{wmap,Ade:2013ktc} and weak-lensing observations \cite{sdss,weaklensing} (for a recent review see \cite{Taoso:2007qk}). Both inflation and dark matter are essential features in the modern cosmological paradigm and cannot be accounted for within the present framework of the Standard Model of particle physics. It is therefore interesting to consider the possibility that the same field accounts for both accelerated expansion in the early universe and the hidden matter component at late times. 

Scalar fields have the interesting property of mimicking fluids with different equations of state depending on the kinematical regime considered. For a homogeneous scalar field $\phi$ with potential $V(\phi)$, we have:
\begin{eqnarray} \label{inflatn_eos}
	\rho_\phi={1 \over 2}\dot\phi^2+V(\phi)~, \qquad p_\phi= {1\over2}\dot\phi^2-V(\phi)~.
\end{eqnarray}
Hence, on the one hand a slowly-varying field, $\dot\phi^2/2\ll V(\phi)$ acts as an effective cosmological constant, which is the regime typically
considered in canonical inflationary models. On the other hand, a field oscillating about the minimum of its potential where $V(\phi)\simeq
m_\phi^2\phi^2/2$ behaves as non-relativistic matter, with $\langle\dot\phi^2/2\rangle= -\langle V(\phi)\rangle$ such that $p_\phi\ll \rho_\phi$
\cite{Turner:1983he}. These two regimes will generically be present in inflationary potentials, which further suggests a common framework for
inflation and dark matter. The main difficulty in achieving such a unified description lies, however, in the fact that inflation must end with a
transition to a radiation-dominated universe, in order to recover the standard ``Big Bang" evolution at least before the freeze-out of light nuclear
abundances takes place \cite{Liddle:2006qz}. An efficient transfer of energy between the inflaton field and radiation generically requires the former
to decay into light degrees of freedom following the period of inflationary slow-roll \cite{reheating}, even though other non-perturbative
processes such as parametric resonance amplification could contribute significantly to the reheating process \cite{Kofman:1997yn}. A plausible
alternative is also that non-equilibrium dissipative processes continuously source a radiation bath during inflation that smoothly takes over as the
field exits the slow-roll regime, a possibility generically known as warm inflation \cite{Berera:1995wh,Berera:2008ar}.

Nevertheless, efficient reheating does not imply the complete decay of the inflaton field and the possibility that a stable remnant is left after
inflation has been considered in a few studies in the literature \cite{thermal,th_infl,kinetic,regeneration,singlet}. In this work, we propose a concrete realization of this generic idea in quantum field theory, where the decay of the inflaton is truly incomplete, occurring
only for a finite period after the end of the slow-roll regime. 

We describe a simple mechanism based on standard Yukawa coupling between the inflaton and fermion fields (and potentially their superpartners), endowed with an appropriate symmetry that protects the full decay of the inflaton field, and discuss the parametric regimes where inflaton decay into such fermions is incomplete. We also discuss the embedding of this generic mechanism in concrete inflationary models and possible scenarios for the interactions between the fermion fields and the Standard Model degrees of freedom that allow for the presence of the latter in the post-inflationary thermal bath. We also show that the inflaton remnant is not necessarily in the form of a coherent condensate of bosonic particles and that, in particular, for not too small couplings the evaporation of this condensate is inevitable. In this case, the thermalized inflaton particles eventually decouple from the radiation bath and their abundance freezes out, yielding a thermal inflaton relic with properties similar to other WIMP 
candidates \cite{Steigman:1984ac}. We suggestively denote this as the ``WIMPlaton" scenario.

In this work, we consider two dynamically distinct scenarios. Firstly, we will consider a minimal model with a single dynamical scalar field that simultaneously drives inflation, reheats the universe through incomplete decay and leaves a stable non-relativistic remnant. Secondly, we consider a (supersymmetric) hybrid inflation model with an additional dynamical waterfall sector, which is responsible for ending inflation and reheating of the universe. In this case, we show that despite reheating being ensured by a different field, the inflaton field must decay in order for radiation to fully come to dominate the energy balance in the universe. As in the minimal model, this decay may nevertheless be incomplete and we discuss the parametric regimes in which the inflaton remnant constitutes a suitable dark matter candidate in the hybrid framework.

This work is organized as follows. In the next section we discuss the minimal model of inflaton dark matter from incomplete decay, starting with its basic properties and dynamics and then discussing the possibility of condensate evaporation, its embedding in a consistent inflationary model and different phenomenological possibilities for reheating the Standard Model degrees of freedom. In Section III we discuss these aspects in the hybrid realization of inflaton dark matter. Finally, in Section IV we summarize our main results and conclusions.


\section{Minimal model}

\subsection{Basic properties and dynamics}

The minimal model for inflaton dark matter considers a single dynamical (real) scalar field, the inflaton $\phi$, with a potential energy $V(\phi)$
such that a period of slow-roll can occur for some field range. We take the inflaton field to be coupled to fermion fields $\psi_+$ and $\psi_-$
through standard Yukawa terms and impose a discrete symmetry\footnote{This subgroup contains only elements that transform simultaneously under $
\mathbb{Z}_2$ and $S_2$.} \cite{wukitung} $C_2\subset \mathbb{Z}_2\times S_2$  on the Lagrangian such that the scalar inflaton transforms under the
$\mathbb{Z}_2$ group as $\phi\rightarrow -\phi$ and the fermions are simultaneously interchanged by the permutation symmetry $\psi_+\leftrightarrow
\psi_-$. This yields for the resulting Lagrangian density:
\begin{eqnarray} \label{Lagrangian}
\mathcal{L}&=& {1\over2}\partial^\mu\phi\partial_\mu\phi-V(|\phi|)\nonumber\\
&+& \bar\psi_+(i\gamma^\mu\partial_\mu-m_f)\psi_+ +  \bar\psi_-(i\gamma^\mu\partial_\mu-m_f)\psi_-\nonumber\\
&-&h\phi\bar\psi_+\psi_+ + h\phi\bar\psi_-\psi_-~,
\end{eqnarray}
where we note that, as a result of the discrete symmetry, the two fermions have the same tree-level mass $m_f$ but opposite Yukawa coupling to the inflaton field. The action of the discrete symmetry is restricted to the inflaton-fermion sector, such that all other fields, including the Standard Model fields, are invariant under this symmetry. 

This symmetry implies that the inflaton potential is an even function of the field and we assume that it is unbroken at the minimum of the potential, which must therefore lie at the origin, $\phi=0$. As a result, since no other terms linear in $\phi$ except for the above Yukawa terms are allowed by the discrete symmetry, the only possible decay channels of the inflaton at the minimum are $\phi\rightarrow \bar\psi_\pm\psi_\pm$. If the inflaton mass at the minimum is given by $m_\phi^2=V''(0)>0$ and we require $m_f>m_\phi/2$, these decays will be kinematically forbidden for $\phi=0$ and the inflaton will be stable when it settles at the minimum of its potential. Hence, the discrete symmetry and the mass hierarchy ensure the stability of the inflaton at late times, such that it will contribute to the present dark matter abundance.

The two fermion fields are indistinguishable apart from their coupling to the inflaton, which leads to a mass splitting away from the origin:
\begin{eqnarray} \label{fermion_mass_splitting}
m_\pm=|m_f \pm h\phi|~.
\end{eqnarray}
This implies that, although inflaton decay into these fermions is forbidden at late times as the field approaches the origin, that need not be the case if the amplitude of field oscillations after inflation is sufficiently large. In particular, decay will be allowed for field values satisfying:
\begin{eqnarray} \label{decay_allowed}
|m_f \pm h\phi|< m_\phi/2~.
\end{eqnarray}
This implies, in particular, that decay is kinematically allowed for field amplitudes $|\phi|\gtrsim m_f/h$. The partial decay widths associated to the two fermionic decay channels are then given by:
\begin{eqnarray} \label{decay_width}
\Gamma_\pm={h^2\over8\pi}m_\phi\left(1-{4m_\pm^2\over m_\phi^2}\right)^{3/2}~,
\end{eqnarray}
with $\Gamma_\phi=\Gamma_+ + \Gamma_-$. Note that due to the opposite sign of the Yukawa couplings, the inflaton will alternately decay into each fermion species as it oscillates between negative and positive values. The inflaton equation of motion as it oscillates about the minimum of its potential is then given by:
\begin{eqnarray} \label{inflaton_equation}
\ddot\phi + 3H\dot\phi + \Gamma_\phi \dot\phi+ m_\phi^2\phi =0~,
\end{eqnarray}
which, upon multiplying by $\dot\phi$ is equivalent to:
\begin{eqnarray} \label{inflaton_density_equation}
\dot\rho_\phi +3H(\rho_\phi+p_\phi)=-\Gamma_\phi \dot\phi^2~.
\end{eqnarray}
The term on the right hand side gives the rate at which energy density is transferred from the inflaton field into the fermions $\psi_\pm$. Let us assume that the fermions quickly thermalize, an assumption that we will check {\it a posteriori}, in the process exciting $g_*$ relativistic degrees of freedom and forming a radiation bath at temperature $T$, with energy density $\rho_R= (\pi^2/30)g_*T^4$. Note that since the fermion masses oscillate due to the varying inflaton field, they only contribute periodically to the number of relativistic degrees of freedom. These may also include the inflaton for $T\gtrsim m_\phi$ and other species such as the Standard Model particles. Note that the latter must be excited before the cosmological synthesis of light nuclear elements takes place, as we discuss in more detail in section IID. For simplicity, we consider a fixed value of $g_*$, which is not a bad approximation since, as we will show later on, our results exhibit only a mild dependence on this parameter.

Energy conservation then implies that the energy lost by the inflaton field in Eq.~(\ref{inflaton_density_equation}) is gained by the radiation bath, which then follows the dynamical equation:
\begin{eqnarray} \label{radiation_equation}
\dot\rho_R+4H\rho_R= \Gamma_\phi \dot\phi^2~.
\end{eqnarray}
Since both the inflaton and the radiation contribute to the energy density in the universe, we may write the Friedmann equation as:
\begin{eqnarray} \label{Friedmann_equation}
H^2={\rho_\phi+\rho_R\over 3M_P^2}={{1\over2}\dot\phi^2+{1\over2}m_\phi^2\phi^2+\rho_R\over 3M_P^2}~.
\end{eqnarray}

\begin{figure}[htbp] 
\centering\includegraphics[scale=1]{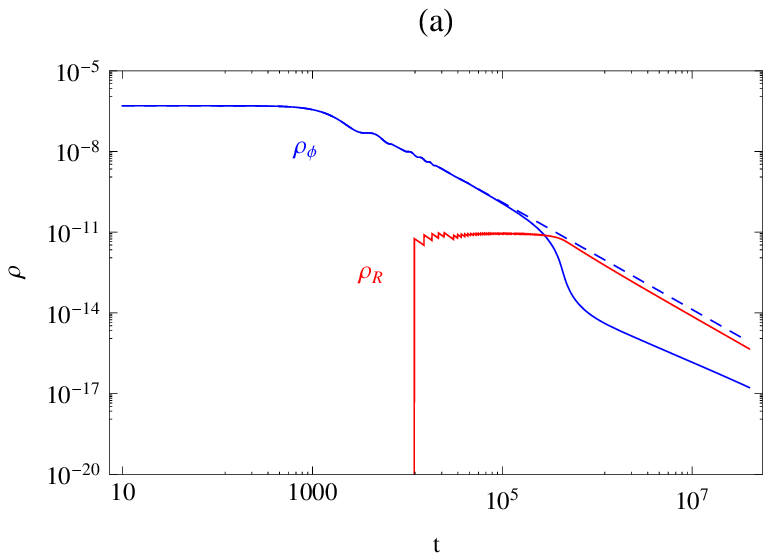}\vspace{0.5cm}
\centering\includegraphics[scale=1]{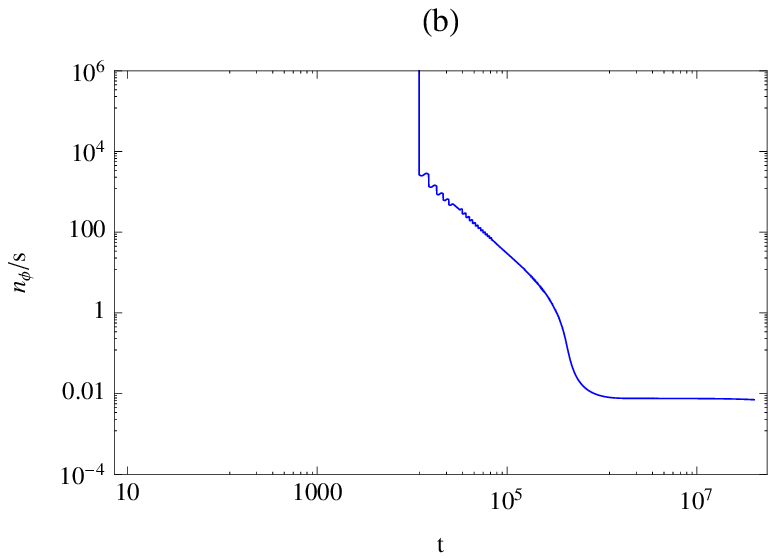}\vspace{0.5cm}
\centering\includegraphics[scale=1]{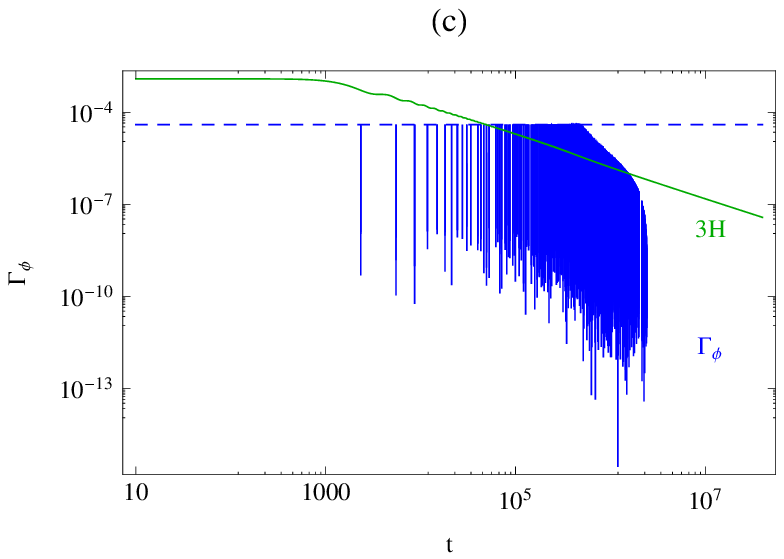}
\caption{Results of the numerical integration of the inflaton-radiation dynamical equations for $m_\phi=10^{-3}M_P$, $\delta=0.1$ and $h=1$, showing the time evolution of (a) the inflaton (solid blue curve) and radiation (solid red curve) energy densities; (b) the inflaton-to-entropy ratio; and (c) the inflaton decay width (solid blue curve) compared to the Hubble parameter (solid green curve). The blue dashed curves in (a) and (c) give the evolution of the inflaton energy density in the absence of decay and the maximum value of the decay width, respectively. All quantities are given in Planck units such that $M_P=1$.}
\label{minimal_sim}
\end{figure}

Eqs.~(\ref{inflaton_equation}), (\ref{radiation_equation}) and (\ref{Friedmann_equation}) then form a complete set of differential equations that can be solved for given choices of the parameters $(m_\phi, m_f, h)$ and initial conditions. The non-linearity of the equations precludes a complete analytical description of the problem, although as we discuss below a few simple analytical estimates can be performed, but it is straightforward to integrate them numerically. It is useful to write the fermion tree-level mass as:
\begin{eqnarray} \label{delta}
m_f={m_\phi\over 2} (1+\delta)~,
\end{eqnarray}
such that the decay is kinematically forbidden (allowed) at late times for $\delta>0$ ($<0$).

Figure \ref{minimal_sim} shows an example of the results obtained by numerically solving the inflaton-radiation equations for $m_\phi=10^{-3}M_P$, $\delta=0.1$ and $h=1$. In this example we take as initial conditions at $t=t_i=10^{-2}m_\phi^{-1}$, $\phi(t_i)=M_P$, $\dot\phi(t_i)=0$ and $\rho_R(t_i)=10^{-16}\rho_\phi(t_i)$, although the results are essentially unchanged as long as the field value is sufficiently large before oscillations begin for the decay to be kinematically allowed as discussed above and the radiation energy density is initially negligible, as should be expected after 50-60 e-folds of inflation.

The plot in Figure \ref{minimal_sim} (a) illustrates the main dynamical features that are generically obtained. The inflaton field begins to oscillate about the origin with frequency $m_\phi$ for $t\sim m_\phi^{-1}$ and behaves initially as cold dark matter, $\rho_\phi\propto a^{-3}$, being the dominant energy component such that $a\propto t^{2/3}$ and $H=2/3t$. While decay is blocked before the onset of oscillations, since $m_\pm\sim h|\phi|\gg m_\phi$, it becomes kinematically allowed as soon as the field goes through the origin. Since in this example the amplitude of the field oscillations largely exceeds the tree-level masses, with $h\Phi\gg m_f$, decay occurs for two narrow field ranges close to (and on both sides of) the origin. The decay width then corresponds to the series of periodic narrow peaks shown in Figure \ref{minimal_sim} (c), with maximum value $\Gamma_\phi^{max}=(h^2/8\pi) m_\phi$. While initially $\Gamma_\phi^{max} \ll 3H$ as illustrated in Figure \ref{minimal_sim} (c), such that the 
inflaton's energy density remains essentially unaffected by the decay into fermions, the source term in the radiation equation quickly becomes significant, leading to a jump in the value of the radiation energy density. The latter remains approximately constant until the inflaton's energy density is sufficiently redshifted. When they become comparable in magnitude, the inflaton effectively decays and radiation takes over as the dominant component. As the field amplitude decreases, the maximum decay width becomes progressively smaller until decay is finally blocked. The oscillating inflaton then becomes stable and once more behaves as cold dark matter, eventually taking over the radiation as the dominant component at later times.

A peculiar dynamical feature of the evolution is the approximate constancy of the radiation energy density achieved just after the first few oscillations. This is inherent to the fact that inflaton decay occurs in short bursts in each oscillation, which does not occur if the decay were allowed for all field values. Before the effect of decay into fermions becomes significant, for $t\gtrsim m_\phi^{-1}$ the inflaton behaves as a damped harmonic oscillator with
\begin{eqnarray} \label{oscillating_inflaton}
\phi(t)\simeq \Phi(t)\sin(m_\phi t+ \alpha_\phi)~, \qquad \Phi(t)=\sqrt{8\over3}{M_P\over m_\phi}{1\over t}~,\nonumber\\
\end{eqnarray}
where $\alpha_\phi$ is a phase depending on the initial field and velocity values. For $h\Phi\gg m_f$, one can easily see that decay into each fermion is allowed during a short period $\tau_d\simeq (h\Phi)^{-1} \ll 2\pi/m_\phi$ as the field goes through the origin, which occurs twice every oscillation period. Since the average decay width in this interval can be taken as $\Gamma_\phi^{max}/2$ and the field velocity is $\dot\phi\simeq m_\phi \Phi$, every half period the radiation energy density increases due to inflaton decay by an amount:
\begin{eqnarray} \label{delta_rho_decay}
\Delta \rho_R^{decay}\simeq {\Gamma_\phi^{max}\over 2} (m_\phi \Phi)^2 (2\tau_d)\simeq {h\over 8\pi} m_\phi^3 \Phi~,
\end{eqnarray}
where we have taken into account the decays into both $\psi_+$ and $\psi_-$. When decay is forbidden, radiation simply redshifts with expansion, which counteracts the enhancement due to decay by an amount:
\begin{eqnarray} \label{delta_rho_hubble}
\Delta \rho_R^{Hubble}\simeq -4H\rho_R (\pi/m_\phi)~.
\end{eqnarray}
Since $H=2/3t = (m_\phi/\sqrt{6}M_P)\Phi$, it is easy to see that the amount of radiation produced by inflaton decay can be balanced exactly by Hubble expansion to yield a constant energy density. Equating $\Delta\rho_R^{decay}=-\Delta\rho_R^{Hubble}$ then gives:
\begin{eqnarray} \label{constant_radiation}
\rho_R\simeq {\sqrt{6}h\over 32\pi^2}m_\phi^3 M_P~,
\end{eqnarray}
which is in very good agreement with the numerical simulations. Noting that, from Eq.~(\ref{radiation_equation}), $\dot\rho_R\simeq 0$ implies $4H\rho_R\simeq \Gamma_\phi \dot\phi^2\simeq \Gamma_\phi \rho_\phi$, we see that that $\Gamma_\phi\sim 3H$ for $\rho_R\sim \rho_\phi$, so that as observed numerically the inflaton energy density is only reduced significantly when it becomes comparable to the radiation energy density.

From Eq.~(\ref{constant_radiation}) we can easily determine the associated temperature, which remains approximately constant up to inflaton-radiation equality and thus constitutes the reheating temperature:
\begin{eqnarray} \label{reheating_temperature}
T_R&\simeq&\left({15\sqrt{6}\over16\pi^4}\right)^{1/4}g_*^{-1/4}h^{1/4}\left({m_\phi\over M_P}\right)^{3/4} M_P\nonumber\\
&\simeq & 2.7\times10^6 g_*^{-1/4}h^{1/4} \left({m_\phi\over 1\ \mathrm{TeV}}\right)^{3/4}~\mathrm{GeV}.
\end{eqnarray}
From this we conclude, in particular, that the reheating temperature is generically larger than the inflaton mass for:
\begin{eqnarray} \label{relativistic_inflaton_bound}
 m_\phi< 5.6\times10^{16} {h\over g_*}\ \mathrm{GeV}~.
\end{eqnarray}

After inflaton-radiation equality, the field decays exponentially fast until its amplitude drops sufficiently for decay into fermions to become inefficient. Numerically, we observe that this occurs before the decay becomes kinematically forbidden, for $\Phi\lesssim m_f/h$, corresponding roughly to when the (maximum) decay width becomes less than the Hubble rate. Since afterwards the field stabilizes and behaves as non-relativistic matter, it is useful to compare its number density $n_\phi=\rho_\phi/m_\phi\propto a^{-3}$ with the radiation entropy density:
\begin{eqnarray} \label{entropy_density}
s={2\pi^2\over45}g_* T^3~,
\end{eqnarray}
which redshifts in a similar way up to changes in the number of relativistic degrees of freedom, with $T\propto a^{-1}$. As illustrated in Figure  \ref{minimal_sim} (b), the ratio $n_\phi/s$ stabilizes after decay becomes inefficient, and numerically we obtain the following expression:
\begin{eqnarray} \label{inflaton_entropy_ratio}
{n_\phi\over s}\simeq 0.5  g_*^{-1/4}h^{-3.6}\left({m_\phi/M_P}\right)^{1.02}f^3(\delta)~,
\end{eqnarray}
where for $0<\delta \lesssim 1$:
\begin{eqnarray} \label{function_delta}
f(\delta)=1+4.8\sqrt{\delta}+0.5\delta~.
\end{eqnarray}
If no other processes change the inflaton particle number density in the oscillating field, as we investigate below in more detail, this ratio will remain constant until the present day. Assuming the oscillating inflaton field accounts for all the dark matter in the universe, the present inflaton-to-entropy ratio is given by:
\begin{eqnarray} \label{inflaton_entropy_ratio_present}
{n_{\phi0}\over s_0}= {3\over4}{T_0\over m_\phi} {\Omega_{c0}\over \Omega_{R0}}{g_{*0}\over g_{*s0}}\simeq 10^{-28}{M_P\over m_\phi}~,
\end{eqnarray}
where $T_0=2.73$ K is the present CMB temperature, while $\Omega_{c0}h^2=0.12$ and $\Omega_{R0}h^2\simeq 4.17\times 10^{-5}$ are the present
abundances of cold dark matter and radiation, respectively \cite{Ade:2013ktc}. We have also taken into account the present difference between the number of relativistic degrees of freedom contributing to the radiation and entropy densities, $g_{*0}\simeq 3.36$, $g_{*s0}\simeq3.91$. Equating (\ref{inflaton_entropy_ratio}) and (\ref{inflaton_entropy_ratio_present}), we obtain for the inflaton mass:
\begin{eqnarray} \label{inflaton_mass_condensate}
m_\phi\simeq 72\, g_*^{0.12}h^{1.78}f(\delta)^{1.49}\ \mathrm{TeV}~.
\end{eqnarray}
%
%
%
For fermion masses above the kinematical limit and of the order of the inflaton mass, $0<\delta\lesssim 1$, and taking $g_*=10-100$, we conclude from Eq.~(\ref{reheating_temperature}) that the reheating temperature is above 100 MeV for Yukawa couplings $h\gtrsim 10^{-6}-10^{-5}$, corresponding to inflaton masses $m_\phi\gtrsim 10$ keV. For larger (and arguably more natural) couplings $h\gtrsim 10^{-3}-10^{-2}$, the inflaton may account for the dark matter in the universe for masses in the GeV$-$TeV range, similarly to the mass range obtained for WIMP-like thermal relics. Qualitatively, it is easy to understand the parametric dependence of the required inflaton mass. The (incomplete) decay into fermions is more efficient for larger couplings, which affect both the overall value of the decay width and the effective fermion masses, and lighter fermions, yielding a smaller inflaton-to-entropy ratio at late times and thus allowing for larger inflaton masses to match the present dark matter abundance. On the other 
hand, smaller couplings and heavier fermions lead to a larger inflaton abundance, which may overclose the universe unless the inflaton is sufficiently light.

We note also that there is no gain in considering finely-tuned fermion masses, i.e. $\delta\ll 0$, since even though decay is kinematically allowed for longer as $\delta$ decreases, decay becomes inefficient when $\Gamma_\phi\sim H$, a condition that becomes independent of the tree-level fermion mass in this limit. Although we have restricted the numerical analysis to values of $\delta\lesssim 1$, we expect incomplete decay to be efficient as long as $\Gamma_\phi>H$ for inflaton field values satisfying the kinematical condition in Eq.~(\ref{decay_allowed}). In particular, since the decay width takes its maximum value $\Gamma_\phi^{max}=(h^2/8\pi)m_\phi$ for $h|\phi|=m_f$, i.e.~when the fermions are effectively massless, and $H\sim m_\phi |\phi|/M_P$ during inflaton-domination, we conclude that decay will be efficient for:
\begin{eqnarray} \label{bound_fermion_mass}
m_f\lesssim h^3 M_P~,
\end{eqnarray}
which allows for quite large fermion masses if the Yukawa coupling is not too suppressed.


\subsection{Condensate evaporation: the WIMPlaton scenario}

Although the imposed discrete symmetry protects the inflaton from
decaying into any other particles except for the fermions $\psi_\pm$,
the above analysis neglects the effects of additional interactions
induced by the Yukawa terms in Eq.~(\ref{Lagrangian}) and which may
play an important role as we discuss below.  

The classical inflaton field corresponds to a collective state of
zero-momentum scalar bosons, assuming that no large field
inhomogeneities are formed 
at the end of the slow-roll inflationary regime. 
Inflaton particles in this condensate can interact with the fermions
that result from its decay and, 
in particular, these fermions can scatter some of the bosons out of
the condensate and promote them to higher-momentum states that
  become part of the thermal bath. These correspond to scattering processes
$\psi_\pm\langle\phi\rangle\rightarrow \psi_\pm \phi$, where we denote
by $\langle\phi\rangle$ and $\phi$ scalar particles in the
zero-momentum condensate and in higher-momentum modes, respectively,
and which are mediated through both $s$- and $t$-channel fermion
exchange. Moreover, these processes may occur as soon as the field
begins oscillating and decay into $\psi_\pm$ becomes kinematically
allowed, potentially leading to the evaporation of the condensate and
the transfer of the inflaton particle number into the thermal bath.  

As we have seen earlier, soon after the onset of inflaton
oscillations, the temperature of the thermal bath rises sharply to a
value that remains approximately constant until inflaton-radiation
equality and that corresponds to the reheating temperature in
Eq.~(\ref{reheating_temperature}). In particular, this temperature
exceeds the inflaton and fermion (tree-level) masses in the parameter
space region that yields the present-day dark matter
abundance. Further assuming that local thermal equilibrium is quickly
achieved, as we check below, we may then take the phase-space
distributions for inflaton and fermion species in the thermal bath to
be the relativistic Bose-Einstein and Fermi-Dirac distributions,
respectively.  Taking into account the above scatterings and the
inverse processes, the net condensate evaporation rate is given by \cite{Kolb:1988aj}: 
\begin{eqnarray} \label{evaporation_rate}
\Gamma_{evap}={1\over n_\phi} \int \prod_{i=1}^4 {d^3 {\bf p}_i\over
  (2\pi)^32E_i} (2\pi)^4\delta^4(p_1+p_2-p_3-p_4)\nonumber\\ 
\times|\mathcal{M}|^2\left[f_1f_2(1+f_3)(1-f_4)-f_3f_4(1+f_1)(1-f_2)\right]~,\nonumber\\ 
\end{eqnarray}
where $\mathcal{M}$ is the scattering amplitude for $\langle\phi\rangle(p_1)\psi_\pm (p_2)\leftrightarrow \phi(p_3)\psi_\pm (p_4)$ and $f_i$ the
corresponding phase-space distribution factors. Since the condensate
is inherently characterized by large occupation numbers $f_1\gg 1$, we
obtain to leading order for $T\gg m_\phi, m_\pm$:
\begin{eqnarray} \label{evaporation_rate_2}
	\Gamma_{evap}\simeq {h^4\over {12}\pi^3}\left(1+\log\left({T\over m_\phi}\right)\right)T~,
\end{eqnarray}
where we have taken into account the contribution of both fermion
species. Note that this is only valid when the fermions are
relativistic, while for 
non-relativistic fermions in local thermal equilibrium the evaporation
rate is exponentially suppressed. This means that during the first few 
oscillations before the inflaton amplitude is significantly reduced,
condensate evaporation only occurs during the short periods where
decay is also 
kinematically allowed. As we have concluded above, for the parameter
values yielding 
the present dark matter abundance we obtain $T_R\gtrsim m_\phi, m_f$,
and since $m_\pm\sim m_f$ after reheating the above expression holds
until 
either the inflaton or the fermions become non-relativistic. Since in
the radiation era $H\simeq (\pi/\sqrt{90})g_*^{1/2}T^2/M_P$,
condensate 
evaporation becomes progressively more efficient as the temperature
drops. We may then determine a lower bound on the Yukawa coupling such
that 
condensate evaporation is inefficient for $T\gtrsim m_\phi, m_f$. For
comparable masses, we obtain:
\begin{eqnarray} \label{evaporation_rate_limit}
	\left.{\Gamma_{evap}\over H}\right|_{T=m_\phi}\simeq {h^4\over 4\pi^3} \left(g_* \over 10 \right)^{-1/2}\left({M_P\over
	m_\phi}\right)\lesssim 1~.
\end{eqnarray}
Using Eq,~(\ref{inflaton_mass_condensate}) for the inflaton mass, we
then obtain for $\delta=1$: 
\begin{eqnarray} \label{evaporation_rate_coupling}
h\lesssim 10^{-5}g_*^{0.28}~.
\end{eqnarray}
As we had seen above, this is in tension with the lower bound on the
coupling required for a reheating temperature above 100 MeV, such that
initial conditions for Big Bang Nucleosynthesis (BBN) are already in
place after inflaton decay \cite{bbn}. This conclusion is essentially common to
all fermion mass values $m_f>m_\phi/2$. Although a more detailed
analysis of the Boltzmann equation determining the evolution of the
inflaton condensate may be required, this estimate indicates that in
the physically interesting parameter range, where the condensate could
account for the present dark matter abundance while satisfying the BBN constraint, condensate evaporation is most likely inevitable.  

Since evaporation simply transfers the zero-momentum condensate
particles into excited states, the conclusion above does not imply
that inflaton particles cannot account for dark matter. One can easily
check that all 4-body processes involving relativistic inflaton and
fermion particles, including scatterings and annihilations, occur at a
rate comparable to the evaporation rate obtained in
Eq.~(\ref{evaporation_rate_2}). Hence, evaporation of the condensate
should lead to a bath of fermions and inflaton particles (as well as
other species) in local thermal equilibrium. In fact, this may even
occur just after the onset of inflaton oscillations, for $H\lesssim
m_\phi$, when the temperature reaches an approximately constant value
$T\gtrsim m_\phi, m_f$ as seen above, for $h\lesssim \mathcal{O}(1)$
couplings. The condensate's energy is in this case quickly transferred
to the thermal bath, increasing its temperature to a maximum value: 
\begin{eqnarray} \label{maximum_reheating_temperature}
T_R^{max}&=& \left({90\over \pi^2}\right)^{1/4}g_*^{-1/4}\sqrt{M_P m_\phi}\nonumber\\
&\simeq& 8.5\times10^{10}g_*^{-1/4}\left({m_\phi\over 1\ \mathrm{TeV}}\right)^{1/2}\ \mathrm{GeV}~.
\end{eqnarray}
Note that this was obtained for $H=m_\phi$, which constitutes only an
upper bound since fermion particles are only produced for $H\lesssim
m_\phi$ as shown by our numerical simulations. The true reheating
temperature will then lie between the value obtained in
Eq.~(\ref{reheating_temperature}) and this maximum value if condensate
evaporation occurs before it decays significantly. The reheating
temperature is given by Eq.~(\ref{reheating_temperature}) if
condensate evaporation only occurs in the radiation era, which
corresponds to Yukawa couplings $h\lesssim 10^{-4}$ for
$\delta\lesssim 1$ and $g_*=10-100$. Note that in this parametric
regime thermalization of the fermions in the plasma can only be
efficient if their interactions with other degrees of freedom are
stronger than those induced by the Yukawa terms, while for $h\gtrsim
10^{-4}$ the latter occur sufficiently fast to maintain local thermal
equilibrium, as we analyze in more detail in section IID.  

After reheating, inflaton and fermion particles will be kept in local
thermal equilibrium by annihilation and elastic scattering
processes. Once these become inefficient, the abundance of inflaton
particles will freeze out, as for other conventional WIMP dark matter
candidates. Assuming this occurs when both the inflaton and the
fermions are non-relativistic, the relevant (fermion $t$-channel)
annihilation cross section is given by: 
\begin{eqnarray} \label{sigma_annihilation}
\sigma_{\phi\phi}\simeq {h^4\over 8\pi m_\phi^2}~,
\end{eqnarray}
which is independent of the fermion mass in this limit. Following the
standard calculation for the thermal relic abundance of a decoupled 
non-relativistic species, we obtain for the inflaton mass:
\begin{eqnarray} \label{inflaton_relic_mass}
	m_\phi\simeq 1.4 h^2\left(\Omega_{\phi0}h_0^2\over 0.1\right)^{1/2}\left({g_{*F}\over 10}\right)^{1/4}\left({x_F\over
	25}\right)^{-3/4}~{\mathrm{TeV}}~,\nonumber\\
\end{eqnarray}
where $g_{*F}$ denotes the number of relativistic degrees of freedom
at freeze-out and $x_F=m_\phi/T_F$, with $T_F$ denoting the freeze-out
temperature. This is somewhat smaller than the mass values obtained
assuming the oscillating inflaton condensate survives until the
present day, although in a comparable range and exhibiting a similar
dependence on the Yukawa coupling.  

This then gives us a more realistic dynamical picture of what we
suggestively denote as the ``WIMPlaton" scenario. After inflation, the
scalar inflaton begins oscillating about the minimum of its potential,
decaying into fermions in short bursts every oscillation. These may
thermalize and excite other degrees of freedom in the plasma, and
scatter off the inflaton particles in the condensate, leading to its
evaporation. Both decay and evaporation increase the relative
abundance of radiation and decrease the amplitude of oscillations,
until eventually radiation becomes dominant and inflaton decay is no
longer kinematically allowed. The stable inflaton particles remain in
thermal equilibrium until the temperature drops below their mass and
they decouple from the plasma, their frozen abundance yielding the
inferred dark matter component of our present universe. 


\subsection{Embedding in a consistent inflationary model}

As we have concluded from the analysis above, the inflaton field can
account for dark matter in the universe at late times for masses below
or around the TeV scale in the WIMPlaton scenario, with a similar mass
range obtained assuming the inflaton condensate does not
evaporate. This implies that the inflaton potential cannot be given
solely by a quadratic mass term, since the amplitude of CMB
temperature anisotropies would yield in this case:  
\begin{eqnarray} \label{quadratic_spectrum}
m_\phi\simeq{ \sqrt{6\pi^2\Delta_\mathcal{R}^2}\over N_e}M_P\simeq 1.4\times 10^{13}\left({60\over N_e}\right)\ \mathrm{GeV}~,\nonumber\\
\end{eqnarray}
which for $50-60$ e-folds of slow-roll inflation largely exceeds the TeV scale. However, our analysis assumed only that $m_\phi$ is the inflaton mass as it oscillates about the minimum of the potential at the origin, while the effective inflaton mass can be much larger if slow-roll occurs for significantly larger field values where self-interactions become important. For example, the discrete $\mathbb{Z}_2$ symmetry allows for quartic self-interactions such that:
\begin{eqnarray} \label{quadratic+quartic}
V(\phi)={\lambda\over 4!}\phi^4+{1\over2}m^2\phi^2~,
\end{eqnarray}
with the quartic term dominating for $|\phi|>\sqrt{12/\lambda}
m_\phi$. Since it increases the inflaton mass, the quartic
self-interaction would make the decay into fermions more efficient and
speed up the reheating process. Our analysis is nevertheless valid if
the quartic term is sub-dominant for the field values $|\phi|\sim
m_f/h$ at which decay into fermions occurs, which requires: 
\begin{eqnarray} \label{coupling_bound_quartic}
h\gtrsim \sqrt{\lambda\over 12}\left({m_f\over m_\phi}\right)\simeq 3\times 10^{-8}\left({m_f\over m_\phi}\right)~,
\end{eqnarray}
where we have used $\lambda\simeq 10^{-14}$ as imposed by the COBE
normalization for inflation with a quartic potential. This is easily
satisfied if the fermions are not much heavier than the inflaton given
the more stringent bounds on the Yukawa coupling discussed
earlier. The addition of a quartic term is, however, not sufficient to
produce a consistent model of inflation, since it predicts a too
red-tilted spectrum for curvature perturbations and a tensor-to-scalar
ratio already outside the bounds obtained by Planck \cite{planck} and BICEP2 \cite{Ade:2014xna}.  

A consistent spectrum may be achieved in warm inflation scenarios \cite{Berera:1995wh}, where the coupling
between the inflaton and other sectors plays a significant role during
inflation. For example, renormalizable couplings between the inflaton
and additional fermionic and bosonic species may lead to dissipation
of the inflaton's energy into radiation during the slow-roll period,
leading to an observationally consistent spectrum if inflaton particle
states thermalize with the other relativistic degrees of freedom
\cite{Bartrum:2013fia}. Since this naturally leads to radiation becoming the dominant
component at the end of the slow-roll regime, the post-inflationary
evolution will necessarily differ from the dynamical picture discussed
in this work. It would nevertheless be interesting to investigate
whether the inflaton can act as a stable dark matter candidate at late
times in this context, although this lies outside the scope of the
present discussion. 

Another interesting possibility is the inclusion of a non-minimal
coupling to the gravitational sector, in particular a coupling between
the inflaton and Ricci scalar of the form $\xi \phi^2 R$, which is
compatible with the discrete $\mathbb{Z}_2$ symmetry. The resulting
inflationary scenario yields a perturbation spectrum that smoothly
interpolates between the minimal quartic model and the Starobinsky
model as the non-minimal coupling constant increases. On the one hand,
the latter is characterized by a low tensor-to-scalar ratio and a
spectral index $n_s=0.96-0.97$ in agreement with the Planck results;
on the other hand, a small non-minimal coupling constant is preferred
to obtain a non-negligible tensor-to-scalar ratio . 
We refer the reader to \cite{Salopek:1988qh, Bezrukov:2008dt} for a more detailed
discussion of these scenarios, since here we are mainly interested in
the post-inflationary dynamics. The effect of the non-minimal coupling
on the effective scalar potential in the Einstein frame becomes
negligible for $\xi\phi^2/M_P^2\lesssim 1 $, such that consistency of
our analysis implies  
\begin{eqnarray} \label{coupling_bound_xi}
h\gtrsim \sqrt{\xi} {m_f\over M_P}~,
\end{eqnarray}
which is generically less stringent than
Eq.~(\ref{coupling_bound_quartic}) for masses in the TeV range and
$\xi< 10^{16}$. 

Finally, an important aspect in embedding the interactions in
Eq.~(\ref{Lagrangian}) within a consistent inflationary model is the
fact that the discrete $\mathbb{Z}_2\times S_2$ symmetry does not
protect the scalar potential from radiative corrections. In
particular, the Yukawa interactions induce loop-corrections of the
Coleman-Weinberg form, which for large inflaton field values take the
leading form: 
\begin{eqnarray} \label{CW_fermions}
\Delta V_f\approx - {h^4\phi^4\over 16\pi^2}\left(\log\left(h^2\phi^2\over \mu^2\right)-{3\over2}\right)
\end{eqnarray}
and therefore induce an effective quartic term in the potential. The
effect of this term does not necessarily spoil the predictions of the
non-minimally coupled quartic model, as discussed in \cite{Okada:2010jf}, although
one must ensure that the observed normalization of the perturbation
spectrum is obtained. While for $\xi \ll 1$ the effective quartic
coupling must have approximately the same value as in the minimally
coupled case, which requires $h\lesssim 10^{-3}$, significantly larger
values can be accommodated for large non-minimal couplings.  

Radiative corrections can, however, be significantly reduced in
supersymmetric scenarios, and a supersymmetric version of the model in
Eq.~(\ref{Lagrangian}) with a $C_2\subset\mathbb{Z}_2\times S_2$ symmetry can be
easily obtained by considering a superpotential of the form: 
\begin{eqnarray} \label{superpotential}
W&=&{h\over2}\Phi\left( Y_+^2-Y_-^2\right)+ {m_f\over 2}\left(Y_+^2 + Y_-^2\right)+\nonumber\\
&+&  {m_\phi\over2}\Phi^2+{\lambda\over 2}\Phi^2 Z~, 
\end{eqnarray}
where the inflaton and fermions $\psi_\pm$ are embedded within the
chiral superfields $\Phi$ and $Y_\pm$, respectively, and the auxiliary
superfield $Z$ induces the quartic term in the inflaton potential
(noting that the discrete symmetry forbids cubic inflaton terms in the
superpotential). Supersymmetry then cancels the leading contributions
of scalars and bosons to the 1-loop Coleman-Weinberg potential, which
becomes:  
\begin{eqnarray} \label{CW_SUSY}
\Delta V_{SUSY}\approx {h^2\over 16\pi^2 }\log\left(h^2\phi^2\over \mu^2\right)V(\phi)~.
\end{eqnarray}
This contribution is thus necessarily smaller than the tree-level
potential $V(\phi)\simeq \lambda^2 |\phi|^4/4$ for $h\lesssim 1$,
therefore avoiding the generation of large effective self-interactions
during inflation.  

Besides the Yukawa terms considered so far, the supersymmetric model
also yields scalar interactions between the inflaton and the scalar
partners $y_\pm $ of the fermions $\psi_\pm$, which apart from SUSY
splittings that vanish at the origin have the same mass
$m_\pm=|m_f+h\phi|$. Trilinear terms in the scalar potential also lead
to the decay $\phi\rightarrow y_\pm y_\pm$, with analogous kinematics
and comparable widths to the fermionic decay channels, therefore
yielding a similar incomplete decay of the inflaton as analyzed
above. We note that the incomplete decay dynamics can be fully
described in terms of scalar fields and is therefore not exclusive of
fermion Yukawa couplings, although the required form of the scalar
mass terms is more naturally motivated within a supersymmetric
context. We also note the existence of quartic terms in the scalar
potential which induce the 3-body decay $\phi\rightarrow z y_\pm
y_\pm$, where $z$ is the scalar component of the $Z$ chiral
multiplet. The associated couplings have opposite signs for $y_\pm$
and proportional to $\lambda$, so that they are typically sub-dominant
with respect to the 2-body decays and they are also kinematically
forbidden at late times, so that they do not affect our earlier
conclusions.  


\subsection{Reheating the Standard Model}

We have so far assumed that the fermions (or scalars as discussed above) resulting from inflaton decay thermalize and excite other degrees of freedom, and in particular it is crucial that Standard Model particles are generated in the thermal bath at temperatures above $\sim$100 MeV so that BBN may occur following the standard freeze-out dynamics of light nuclear abundances.

As we have briefly discussed above, the fermions themselves cannot be treated as fully relativistic degrees of freedom before the inflaton decays sufficiently, since their mass varies between small and large values as the inflaton oscillates about the origin. For the short periods when they are light and decay is allowed, thermalization through Yukawa interactions can be quite efficient. For example, fermion-fermion scatterings through inflaton $s-$channel exchange occurs at a rate:
\begin{eqnarray} \label{fermion_thermal_rate}
\Gamma_{\psi\psi}\simeq {9\zeta(3)\over 16\pi^3}h^4 T~.
\end{eqnarray}
Since after the onset of oscillations $H\lesssim m_\phi$ and taking the temperature value obtained in Eq.~(\ref{reheating_temperature}) from inflaton decay, we have:
\begin{eqnarray} \label{fermion_thermal_rate_2}
{\Gamma_{\psi\psi}\over H}>{\Gamma_{\psi\psi}\over m_\phi}\simeq 60 g_*^{-1/4}h^{17/4}\left({m_\phi\over 1\ \mathrm{TeV}}\right)^{-1/4}~,
\end{eqnarray}
such that for $\mathcal{O}(1)$ couplings Yukawa interactions can lead to thermalization from the start of oscillations. Recall that, as seen above, evaporation of the condensate occurs at a comparable rate, so that as the temperature rises due to evaporation both processes become progressively more efficient. If the Yukawa couplings are more suppressed, one may also envisage scenarios where for example the fermions are charged under a gauge symmetry that is unbroken at the relevant temperatures and thermalization occurs through gauge boson exchange for sufficiently strong couplings.

Outside the inflaton field range for which the fermions are effectively light, their number density will become Boltzmann-suppressed, and any interactions will necessarily become inefficient in keeping the fermions in local thermal equilibrium. We thus expect them to decouple for most of the oscillation period, transferring their entropy into light degrees of freedom such as the inflaton itself or e.g.~gauge bosons. If interactions occur faster than the inflaton oscillation rate $m_\phi$, as given in Eq.~(\ref{fermion_thermal_rate_2}) for the Yukawa scattering processes, fermions will drop in and out of local thermal equilibrium as they oscillate between the relativistic and non-relativistic regimes. This will lead to an oscillating $g_*$, but as mentioned at the start of our discussion this is of little consequence since our results exhibit only a mild dependence on this parameter.

Since the inflaton is typically taken as a gauge singlet, the structure of the Yukawa interactions implies that $\psi_\pm$ are non-chiral fermions (either Dirac or Majorana), as opposed to the known SM fermions, being thus unlikely that they are explicitly charged under the SM gauge group. 
There are nevertheless several possibilities for exciting the SM degrees of freedom in the plasma either before or after radiation comes to dominate.

The simplest case is perhaps that of unstable $\psi_\pm$ fermions decaying into a light scalar and a light fermion, for which the decay width is:
\begin{eqnarray} \label{decay_width_fermions}
\Gamma_{\psi_\pm}={h_f^2\over 16\pi}m_\pm~, 
\end{eqnarray}
where $h_f$ denotes the associated coupling constant. Note that this is computed in the fermions' rest frame, whereas in the plasma's frame an additional time dilation factor $m_\pm/T$ suppresses the decay for relativistic fermions. For a significant part of the energy density in radiation to be transferred into these degrees of freedom, the fermions must decay before they become non-relativistic. Thus, requiring $\Gamma_{\psi_\pm}\gtrsim H$ for $T\gtrsim m_f$ in the radiation era (where $m_\pm\simeq m_f$), we obtain the following bound on the coupling:
\begin{eqnarray} \label{coupling_bound_decay}
h_f\gtrsim 8\times 10^{-8}g_{*f}^{1/4}\left({m_f\over 1\ \mathrm{TeV}}\right)^{1/2}~,
\end{eqnarray}
where $g_{*f}$ is the number of relativistic degrees of freedom at $T= m_f$. In addition, for this to happen before BBN we require $m_f\gtrsim 100$ MeV. If the inflaton and fermion masses are comparable, this corresponds to $h\gtrsim 0.01$ according to Eq.~(\ref{inflaton_relic_mass}). These light degrees of freedom may correspond to SM particles if, for example, the fermions coupled to the inflaton field correspond to a pair of degenerate sterile neutrinos, which are singlets under the SM gauge group and may decay into a Higgs-lepton pair through Yukawa terms of the form $h_fH\bar{l}\psi_\pm$. Note that this requires $m_f>m_H=125$ GeV \cite{Aad:2012tfa} and hence $m_\phi<250$ GeV, which is compatible with the present dark matter abundance for couplings $h\lesssim 0.7$ from Eq.~(\ref{inflaton_relic_mass}).

If the fermions are stable, another possibility for reheating the SM degrees of freedom is through efficient annihilation. A possible scenario is for the fermions to be charged under a hidden $U(1)_X$ gauge group, which may mediate fermion scatterings and thus improve the thermalization efficiency. This $U(1)_X$ hidden photon may be kinetically mixed with the SM photon or hypercharge gauge boson $Y^\mu$ through a term of the form $F_X^{\mu\nu}F_{\mu\nu}^Y$, which may be generated radiatively if there are fields charged under both gauge groups or simply via gravitational interactions, as happens e.g.~in string theory. Diagonalization of the gauge kinetic terms then induces a small electric charge for the fermions $\psi_\pm$, such that they may annihilate into SM charged particles via $s$-channel photon exchange, $\psi_\pm\psi_\pm\rightarrow \gamma\rightarrow qq, ll$. At high temperature the annihilation cross section is given by the Thomson scattering formula and the corresponding interaction rate for 
relativistic species is then given by:
\begin{eqnarray} \label{Thomson_scattering}
\Gamma_{th}\simeq {4\zeta(3)\over \pi}\epsilon^2\alpha^2 N_{ch}T~,
\end{eqnarray}
 where $\epsilon$ is the ``mili-charge" of the fermions $\psi_\pm$, $\alpha$ is the fine-structure constant and $N_{ch}$ is the effective number of charged species in the final state, which for the full SM is $N_{ch}=20/3$. In the radiation era, since $H\propto T^2$ annihilation becomes more efficient at smaller temperatures. Once $T\lesssim m_f$, however, the fermion abundance becomes Boltzmann-suppressed and annihilations can no longer be efficient. Thus, requiring that SM species are excited before $\psi_\pm$ become non-relativistic, we obtain the following bound on the mili-charge:
\begin{eqnarray} \label{milicharge_bound_mass}
\epsilon\gtrsim 5\times 10^{-7}g_{*f}^{1/4}\left({N_{ch}\over 20/3}\right)^{-1/2}\left({\alpha^{-1}\over 128}\right)\left({m_f}\over 1~\mathrm{TeV}\right)^{1/2}~.\nonumber\\
\end{eqnarray}
 This bound is not very stringent for fermion masses in the GeV-TeV range, where the main constraints come from direct collider searches (including the LHC) yielding $\epsilon\lesssim 0.1$ for $1\ \mathrm{GeV}\lesssim m_f\lesssim \mathrm{few}\times 100$ GeV and indirect bounds from the CMB anisotropy spectrum, which yield $\epsilon\lesssim 10^{-4}$ for few$\times 100\ \mathrm{GeV}\lesssim m_f\lesssim \mathrm{few}\times $TeV based on the effect of mili-charged particles on the baryon-photon oscillations (for a gauge coupling $g_X=0.1$). Note that for masses above the TeV range, mili-charged particles may give a too large contribution to the dark matter abundance in the universe, and more stringent bounds on $\epsilon$ apply in this case (see \cite{Vogel:2013raa} and references therein). This thus constitutes a promising scenario with potential for experimental probing in the near future.
 
\begin{figure}[htbp]
\centering\includegraphics[scale=0.53]{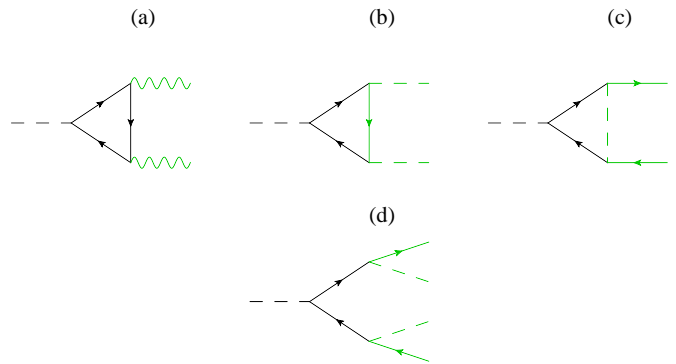}
\caption{Feynman diagrams for the 2-body decay of the inflaton into (a) gauge bosons, (b) light scalars and (c) light fermions, induced at the 1-loop level by gauge and Yukawa interactions of the $\psi_\pm$ fermions. In (d) we also show the 4-body decay of the inflaton induced by the exchange of virtual $\psi_\pm$ modes with Yukawa interactions with other light species. For clarity, all light fields are represented by green lines.}
 \label{decay_diagrams}
\end{figure}

The discrete $C_2\subset\mathbb{Z}_2\times S_2$ symmetry protects the inflaton from decaying at late times, thus constituting a viable dark matter candidate. One can consider, however, scenarios where this symmetry is broken and the inflaton is only meta-stable, with a lifetime larger than the age of the universe, $t_0\sim 14$ Gyrs. Note that interactions between the fermions $\psi_\pm$ and other light degrees of freedom as in the scenarios outlined above can induce the decay of the inflaton through radiative effects or processes mediated by off-shell fermions. A few examples of these processes are illustrated in Figure \ref{decay_diagrams} and in all cases the contribution of $\psi_+$ and $\psi_-$ cancels if the discrete symmetry is exact.
 
As an example, we have considered the case where $\psi_\pm$ are unstable, decaying into a light fermion and scalar, which induce diagrams (b)-(d) in Figure \ref{decay_diagrams}. Since these processes have comparable magnitudes, we have computed the inflaton decay width for the 1-loop process (b), where it decays into two light scalars. For concreteness we consider the case where $\psi_\pm$ are slightly non-degenerate with $m_-=m_f$ and $m_+= m_f(1+\Delta)$, for $\Delta\ll 1$. This gives for $m_\phi=2m_f$ ($\delta=0$):
\begin{eqnarray} \label{inflaton_decay_loop}
\Gamma_\phi^{(b)}={h^2h_f^4\over 64\pi^5}m_\phi\Delta^2~,
\end{eqnarray}
which is more suppressed for larger values of $m_f$.

For $m_\phi$ yielding the correct relic abundance to account for dark matter and using the lower bound on $h_f$ obtained in Eq.~(\ref{coupling_bound_decay}), we obtain the following upper bound on the inflaton lifetime:
\begin{eqnarray} \label{inflaton_lifetime}
\tau_\phi &<& 15.7\ g_{*f}^{-1}\left({m_\phi\over 10\ \mathrm{GeV}}\right)^{-4}\left({\Delta\over 0.01}\right)^{-2}\, \mathrm{Gyrs}\, ,
\end{eqnarray}
where we have taken the reference values for the present dark matter abundance and freeze-out parameters in Eq.~(\ref{inflaton_relic_mass}). Hence, we conclude that significant violations of the discrete symmetry can still yield a sufficiently long-lived inflaton for $100\ \mathrm{MeV}\lesssim m_\phi\lesssim 10$ GeV, such that the fermions $\psi_\pm$ decay before BBN and while they are still relativistic. For heavier inflaton particles, values of $\Delta$ below the percent level are required, signaling that the discrete symmetry most hold to a high degree of accuracy in this regime.

It is thus clear from the examples above that if the inflaton field can only decay to the $\psi_\pm$ fermions for a finite period following the end of the slow-roll regime, becoming (meta-)stable at late times, it may account for the dark matter in the universe while allowing for successful reheating of the SM particles and setting the necessary conditions for BBN.
 

\section{Hybrid model}
\subsection{Basic properties and dynamics}

An alternative framework to the one considered in the previous section is the supersymmetric hybrid model \cite{susyhybrid}. 
In this scenario, the inflaton decay products need not include or interact with the Standard Model degrees of freedom, since the additional waterfall sector can be responsible for reheating after inflation \cite{hybridinf, hybridinf2}. 

For the same reasons exposed in the minimal model, the symmetry $C_{2}\subset\mathbb{Z}_{2}\times S_2$ is imposed on the superfield containing the inflaton and all the superfields that it couples directly to. Hence, for the superpotential to be invariant under the action of this group,  
the inflaton is, as before, coupled to  a pair of superfields $Y_{\pm}$ which contain the fermions $\psi_{\pm}$ with masses $m_f$, and to a waterfall sector with a pair of superfields $X_{\pm}$. The group $C_{2}$ simultaneously changes $\Phi \rightarrow -\Phi$ and interchanges the superfields $Y_+ \leftrightarrow Y_-$ and $X_+\leftrightarrow X_-$. 

The discrete symmetry forbids the linear term in the superpotential that is typically considered in SUSY hybrid inflation to generate the constant vacuum energy driving accelerated expansion. This may nevertheless be generated either by a D-term contribution \cite{Dterm}, or through a non-vanishing F-term coming from a SUSY breaking sector \cite{BasteroGil:1997vn}. In addition, in order to ensure the stability of the inflaton at late times, so that it may account for dark matter, the $C_2$ symmetry must be preserved in the ground state, implying equal vacuum expectation values for the scalar components of both waterfall fields. One possibility to satisfy this condition and simultaneously generate a constant vacuum energy is to introduce an additional ``driving" superfield, $Z$, which is not charged under the discrete symmetry and is coupled to the waterfall sector, along the lines proposed in \cite{Antusch:2009ef}. We thus consider a superpotential of the form:
\begin{eqnarray} \label{hybrid_superpotential}
W &=&{g\over2}\Phi (X_+^2-X_-^2)+{h\over 2}\Phi(Y_+^2-Y_-^2)+{m_f\over2}(Y_+^2+Y_-^2)\nonumber \\
&+&{\kappa\over 2} Z(X_+^2+X_-^2-M^2)\nonumber\\
&+& {h_\chi\over 2}(X_++X_-)Q^2+\ldots~,
\end{eqnarray}
where $M$ is a constant mass scale and we have included a coupling between the waterfall superfields and additional chiral superfields $Q$ which give their decay products. The dots indicate additional terms that may be added, involving the inflaton and the superfield $Z$. In particular, if the scalar component of the latter has a sufficiently large mass, either from superpotential terms, soft masses from SUSY breaking in other sectors or non-minimal terms in the K\"ahler potential, its expectation value will be set to zero both during and after inflation. The global minimum of the scalar potential will then lie along the real direction $\langle X_+\rangle =\langle X_+\rangle \equiv \chi/\sqrt{2}$, which preserves the discrete symmetry, and the scalar potential relevant for the inflationary and post-inflationary dynamics has the usual hybrid form:
\begin{equation}
  V(\phi,\chi)=\frac{\kappa^2}{4}(\chi^2-M^2)^2+\frac{g^2}{2}\phi^2\chi^2+\ldots~,
	\label{scalarpot}
\end{equation} 
where $\phi=\sqrt{2}\langle \Phi\rangle$ is the real inflaton scalar field. We recover the usual SUSY hybrid case for $\kappa=g/\sqrt{2}$ and for simplicity we will consider this parametric regime, although our analysis can be extended to the generic case.

Inflation takes place for amplitudes of the inflaton field larger than a critical value, $\phi>\phi_c=M/\sqrt{2}$, such that the waterfall
field is held at the origin $\chi=0$. As the inflaton rolls towards its minimum at $\phi=0$, its amplitude falls below the critical value and the waterfall field can roll to its true vacuum at $\chi=M$, thus ending inflation. After that point, both fields start to oscillate around its respective minima, triggering the process of reheating the universe into a radiation era. 

In this scenario, the inflaton cannot decay into either the bosonic or fermionic components in the waterfall sector due to kinematical blocking. This is easy to see at the global minimum, where $m_\phi=m_\chi = gM$, but extends to all field values. As in the minimal model, the inflaton can decay into the $Y_\pm$ fields and the decay will be incomplete for $m_f> m_\phi/2 = gM/2$. For simplicity, we assume that the scalar components of the $Y_\pm$ fields acquire large soft masses from SUSY breaking and focus on the fermionic decay channels, noting that the inclusion of both channels will not change our conclusions significantly. 

The waterfall fields will decay into the $Q$ sector fields and, for similar reasons, we include only the fermionic decay channels in this case as well. We assume that these fields are light, eventually leading to the complete decay of the waterfall sector and reheating the universe. Note that neither the inflaton nor the waterfall field can be completely stable in order to reheat the universe,  since they carry a comparable amount of the energy density after inflation. In particular, if the inflaton were completely stable and behaved as dark matter at all times, the decay of the waterfall field would only convert half of the total energy density into radiation. Its incomplete decay will then reduce the inflaton abundance and hence allow for an efficient reheating once the waterfall field decays. For the inflaton to decay incompletely before the waterfall field, we require $h\gtrsim h_\chi$, which is the parametric regime on which we will focus henceforth.

The evolution equations driving the post-inflationary dynamics of the inflaton and waterfall fields, as well as their decay products which we assume to quickly thermalize, are then given by:
\begin{eqnarray}
& & \ddot{\phi}+3H\dot{\phi} + g^2\chi^2\phi=-\Gamma_\phi\dot{\phi},\label{peq}\\
& & \ddot{\chi} + 3H\dot{\chi} + \frac{g^2}{2}(\chi^2 - M^2 + 2 \phi^2)\chi=-\Gamma_\chi\dot{\chi},\label{weq}\\
& & \dot{\rho}_R + 4H\rho_R = \Gamma_\chi\dot{\chi}^2+\Gamma_\phi\dot{\phi}^2,\label{req}
\label{eom}
\end{eqnarray}
where  the decay width of the inflaton is $\Gamma_\phi=\Gamma_+ + \Gamma_-$, with $\Gamma_\pm$ given by Eq. (\ref{decay_width}), while the decay width of the waterfall field is given by:
\begin{equation}
	\Gamma_{\chi}=\frac{h_\chi^2 m_\chi}{8\pi}~,	
	\label{decay_chi}
\end{equation}
where we neglect the $Q$ fermion masses, which we have checked numerically to be a good approximation for couplings $h_\chi \lesssim 0.1$ in the parameter range of interest to our discussion.

\begin{figure}[htbp] 
\centering\includegraphics[width=\columnwidth]{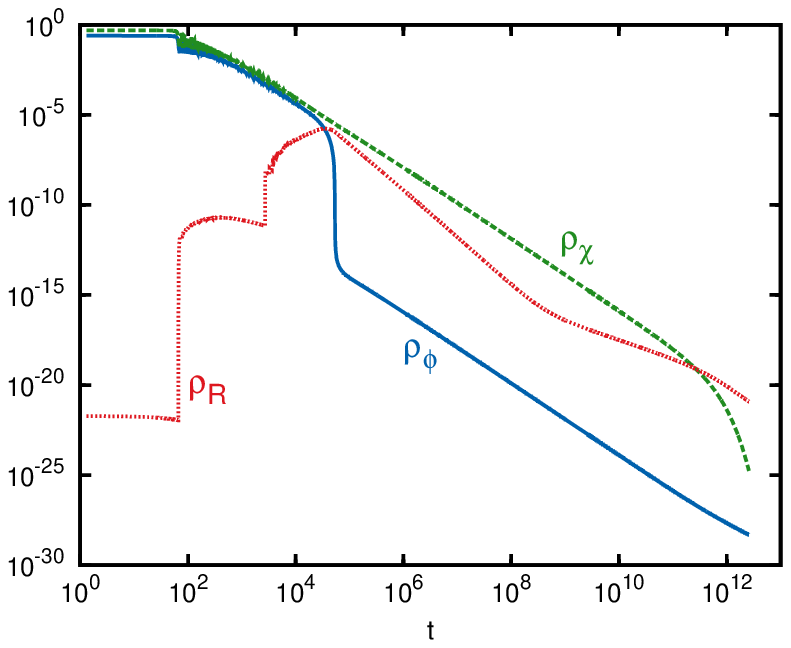}\vspace{0.5cm}
\centering\includegraphics[width=\columnwidth]{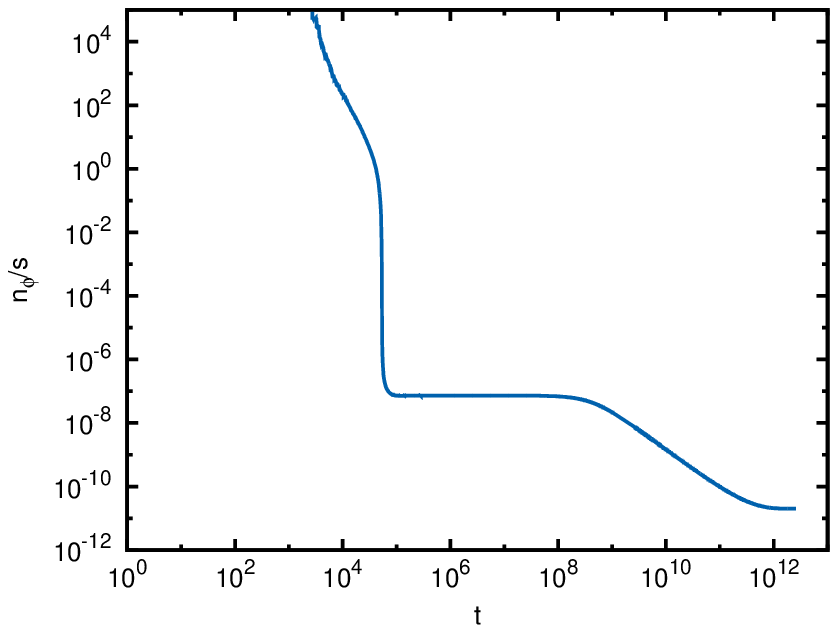}\vspace{0.5cm}
\caption{Results of the numerical integration of the inflaton-waterfall-radiation dynamical equations for $M=10^{-2}M_P$,
  $g=10^{-5}$, $h=1$, $h_{\chi}=10^{-5}$ and $\delta=0.02$, showing the
  time evolution of (a) the inflaton (solid blue curve), the waterfall
  (dashed green curve) and radiation (dotted red curve) energy
  densities; (b) the inflaton-to-entropy ratio. All quantities are
  given in Planck units such that $M_P=1$.} 
\label{hybrid_evol}
\end{figure}

Eqs.~(\ref{peq}),(\ref{weq}) and (\ref{req}), together with the Friedmann equation,
\begin{equation}
H^2=\frac{\rho_\phi+\rho_\chi
  +\rho_R}{3M_P^2}={{1\over2}\dot\phi^2+{1\over2}\dot\chi^2+
  V(\phi,\chi)+\rho_R\over 3M_P^2}~, 
\label{friedmann_hybrid}
\end{equation}
form a complete set of differential equations that can be solved numerically given a set of parameters $(g,M,m_f,h,h_\chi)$ and initial conditions.  Fig. \ref{hybrid_evol} shows an example of the numerical solution for the parameter values $M=10^{-2}M_P$, $g=10^{-5}$, $h=1$, $h_{\chi}=10^{-5}$ and $\delta=0.02$, the latter being defined in Eq.~(\ref{delta}). In this example we have considered initial conditions such that the fields are close to their values at the end of inflation and have small velocities, $\phi(0)=1.0001\phi_c$, $\chi(0)=0$, $\dot{\phi}(0)=\dot{\chi}(0)=-10^{-4}gM^2$, with also $\rho_R(0)=0$ since any pre-inflationary radiation should be exponentially diluted by the accelerated expansion. However, we have checked numerically that the results do not show a strong dependence on the choice of initial conditions. 

At the beginning of the evolution, both the inflaton and the waterfall field mimic a pair of coupled matter fluids which give a roughly equal contribution to the total energy density. After the incomplete decay of the inflaton becomes efficient, its energy density is transferred into the radiation bath, while the waterfall field evolves as an effective non-interacting matter field in an expanding universe. 

Since radiation is diluted more quickly than matter by the cosmological expansion, the universe experiences an era of matter domination until the waterfall field effectively decays and reheats the universe. We then enter the standard radiation-dominated era of Big Bang cosmology, with the oscillating inflaton field remnant behaving as a cold dark matter component. 

The temperature after the effective decay of the inflaton, $T_D$, can be determined analytically following the same reasoning used in minimal
model to compute the reheating temperature, giving:
\begin{equation}
	T_D=\left(\frac{15\sqrt{3}}{16\pi^4}\right)^{1/4}g_{*D}^{-1/4}h^{1/4}\left(\frac{m_\phi}{M_P}\right)^{3/4}M_P, 
	\label{temp_decay}
\end{equation}
where $g_{*D}$ is the effective number of light degrees of freedom when the inflaton effectively decays. Unlike in the minimal model, the reheating temperature is not controlled by the inflaton but rather by the waterfall decay, corresponding to the temperature for which $\Gamma_{\chi}=H$:
\begin{equation}
	T_R \simeq 0.23 g_{*R}^{-1/4}h_{\chi}\left(m_\phi M_P\right)^{1/2}~,
\end{equation}
with $g_{*R}$ being the effective number of light degrees of freedom at reheating and where we used $m_\phi=m_\chi$ since the fields are close to the global minimum at this stage.

\begin{figure}[htbp] 
\centering\includegraphics[width=\columnwidth]{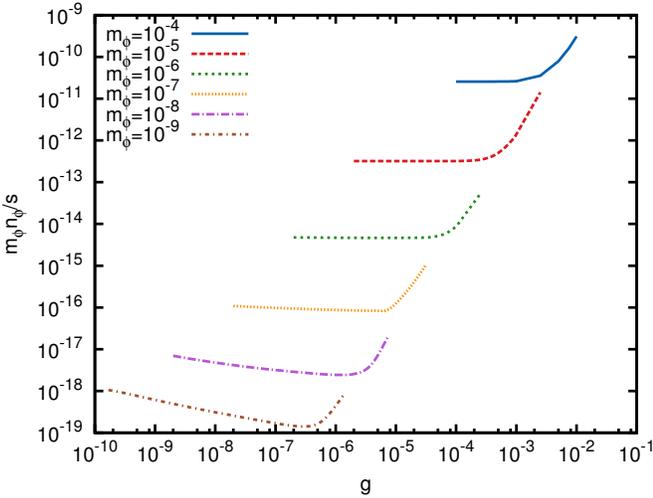}\vspace{0.5cm}
\caption{Dependence of the inflaton-to-entropy ratio after reheating
  multiplied by the inflaton mass in Planck units for different values
  of the parameters in the potential, $g$, $M$ for $\delta=0.02$,
  $h=1$ and $h_{\chi}=10^{-5}$.} 
\label{g_dependence}
\end{figure}

The computation of the inflaton dark matter abundance after reheating is more involved than in the minimal scenario due to the
larger set of parameters in the hybrid scenario and the coupling between the inflaton and waterfall fields. The effect of the parameters of the
potential in Eq.~(\ref{scalarpot}), $g$ and $M=m_\phi/g$ on the final abundance is shown in Fig. \ref{g_dependence}. 

For large values of the coupling $g$, the inflaton-to-entropy ratio grows with $g$ since the system is strongly coupled and the waterfall field
transfers a significant part of its energy to the inflaton, therefore increasing its abundance. For very low values of the coupling, on the other hand, the inflaton-to-entropy ratio shows a very mild dependence on this coupling. In this case the Hubble parameter $H\sim m_\phi^2/g$ is very large, so that the condition for the incomplete decay to be efficient, $\Gamma_\phi>H$, cannot be maintained for sufficiently long and the abundance is not so drastically reduced. Numerically, we take the endpoint of the curves at small $g$ values such that $V_0^{1/4}= \sqrt{g}M = m_\phi/\sqrt{g}= 10^{16}$ GeV, of the order of the GUT scale, which gives roughly the upper bound on the scale of inflation. The vacuum energy scale therefore diminishes when moving from left to right along each curve of constant $m_\phi$ in Fig. \ref{g_dependence}.

We will restrict the remainder of our analysis to the values of $g$ for which the incomplete decay exhibits its maximal efficiency, therefore yielding the lowest inflaton abundance for a given mass $m_\phi$. Numerically, we find these values to be given by:
\begin{equation} 
	g\simeq 0.04\left(\frac{m_\phi}{M_P}\right)^{0.57}\delta^{-0.08}.
	\label{g_minimum}
\end{equation}

With this relation, we obtain for the inflaton-to-entropy ratio from the numerical results:
\bea
	\frac{n_\phi}{s}&\simeq & 134 g_{*}^{-1/4}h_\chi h^{-3.42
          \beta(h)}\delta^{2 \gamma(\delta)}\left(\frac{m_\phi}{M_P}\right)^{0.7}\! \!, 
	\label{ns_hybrid} \\
\beta(h) &=& 1-\frac{1}{6}\log_{10}h \,, \\
\gamma(\delta) &=& 1+0.115\log_{10}\delta \,. 
\eea
This expression is equivalent to Eq. (\ref{inflaton_entropy_ratio}) in the minimal model, with the different powers on the parameters reflecting
the more complicated dynamics present in this scenario. Equating (\ref{inflaton_entropy_ratio_present}) and (\ref{ns_hybrid}), we then obtain for the inflaton mass yielding the observed dark matter abundance:
\begin{equation}
m_\phi \simeq 442~g_{*}^{0.15} \left(\frac{h_{\chi}}{10^{-3}} \right)^{-0.6}h^{-0.6 \beta(h)}\delta^{-0.6 \gamma(\delta)}\, \mathrm{GeV}\,.
	\label{inflaton_mass_condensate_hybrid}
\end{equation}
For couplings $h\sim \mathcal{O}(1)$ and $h_\chi\sim 10^{-3}-10^{-5}$, the inflaton may account for the dark matter in the universe with masses in the GeV-TeV range, similarly to the minimal model, while predicting a reheating temperature well above the BBN constraint. The qualitative dependence of the inflaton mass on the Yukawa coupling $h$ can be understood using the same arguments as in the minimal realization for inflaton dark matter described earlier in this work. 

The parameter $h_\chi$ determines the duration of the reheating process, which is due to the decay of the waterfall field.  During that process, there is entropy production \cite{turner} and as a consequence the abundance $n_\phi/s$ will be further reduced, as shown in Fig. \ref{hybrid_evol} (b). The dilution factor is given by $\Upsilon = S_{eq}/S_R$, with $S$ being the entropy and the subscript ``$eq$'' denoting the time at which the $\chi$ thermalized decay products start dominating the radiation bath, at a temperature $T_{eq} > T_R$. The dilution factor is then given by:  
\be
\Upsilon=\left(\frac{T_R}{T_{eq}} \right)^5 \simeq \frac{5}{3}
\left(\frac{T_R}{T_D} \right) \left(\frac{g_{*D}}{g_{*R}} \right) 
\,,
\ee
with $T_D$ given in Eq. \ref{temp_decay}. The smaller the coupling
$h_\chi$, the longer the reheating process 
and the smaller $T_R$, and the more efficient the reduction of 
the abundance. Therefore larger inflaton masses are allowed to match
the present dark matter abundance.


\subsection{Condensate evaporation: the WIMPlaton scenario} 

\begin{figure}[t] 
\centering\includegraphics[width=\columnwidth]{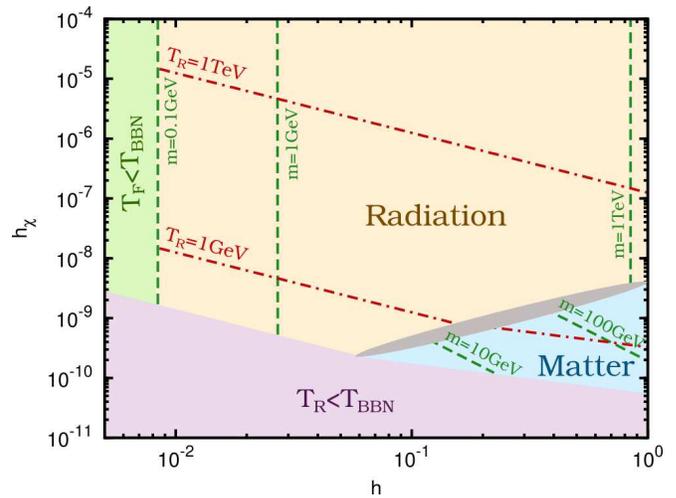}\vspace{0.5cm}
\caption{Parameter space of the hybrid model of inflaton dark matter
  with $\delta=1$. In the orange (blue) region the abundance of the
  inflaton accounts for the present dark matter energy density and the
  freeze-out occurs in the radiation (waterfall=matter) era. The purple
  (green) region is excluded because the reheating (freeze-out)
  temperature is below $100$ MeV. Dashed (dashed-dot) lines are curves
  of constant inflaton mass (reheating temperature). The grey area
  represents the transition between the regions where the freeze-out
  takes place in the radiation and waterfall era.} 
\label{space}
\end{figure}

The effect of additional processes induced by the Yukawa coupling $h$ in the evaporation of the condensate of zero-momentum inflaton
particles is essentially the same that has been described for the minimal model, with the rate of evaporation given by Eq.~(\ref{evaporation_rate_2}). However, the possibility of the fermions becoming non-relativistic during the matter era, where the dominant contribution to the energy density comes from the waterfall field, gives a different lower bound on the Yukawa couplings such that evaporation is inefficient for $T\gtrsim m_\phi, m_f$: 
\begin{equation}
	h_\chi \lesssim 5\times 10^{-10}h^{-\frac{3}{16}\log_{10}{h}}\left(\frac{g_*}{10}\right)^{13/32}.
\end{equation}
Nevertheless, as we found in the minimal scenario of inflaton dark matter, this bound is in tension with the lower bound on the couplings
consistent with a reheating temperature above $100$ MeV. Therefore, in the region where the condensate may account for the present dark
matter and the reheating temperature is consistent with the BBN constraint, condensate evaporation is most likely inevitable. The evaporation process produces a bath of fermions and inflaton particles kept in local thermal equilibrium by annihilation and elastic scatterings that eventually become inefficient, at which point the inflaton abundance freezes out as in the standard WIMP scenario. 

If freeze-out occurs after the waterfall field has decayed, i.e.~in the radiation-dominated era, we obtain the same value for the inflaton mass as in the minimal model, given in Eq. (\ref{inflaton_relic_mass}). Freeze-out takes place in the waterfall-dominated matter for values of  $h_\chi$ satisfying:
\bea
h_\chi\lesssim  4\times 10^{-9} h\left(\frac{g_{*F}}{10}\right)^{3/8}\left(\frac{x_F}{25}\right)^{-11/8}\left(\frac{\Omega_{\phi
            0}h_0^2}{0.1}\right)^{1/4}~,\nonumber\\
\eea
which then yields an upper bound on the WIMPlaton mass:
\bea            
m_\phi &\lesssim& 756
h^2\left(\frac{g_{*F}}{10}\right)^{1/4}\left(\frac{x_F}{25}\right)^{-3/4}\left(\frac{\Omega_{\phi
    0}h_0^2}{0.1}\right)^{1/2}\mathrm{GeV}~,\nonumber\\
\eea
where the annihilation cross section is given by Eq. (\ref{sigma_annihilation}). In Fig. \ref{space} we summarize on the plane $(h,h_{\chi})$ the different possibilities for inflaton dark matter in the hybrid model. In both regions where the freeze-out of the inflaton abundance occurs either in the radiation or waterfall era, the inflaton can account for the present dark matter abundance for masses in the GeV-TeV range with the reheating and freeze-out temperatures being well above the limit imposed by BBN. This shows that the WIMPlaton scenario introduced earlier is not an exclusive feature of the minimal model, with a single dynamical field, but also occurs in other models of inflation with additional dynamical fields.

While the inflaton mass values corresponding to the observed dark matter abundance are not very different in the two realizations that we have analyzed, in the hybrid scenario there are novel phenomenological possibilities. In particular, the inflaton decay products need not interact with the SM degrees of freedom, since these may be excited only after the decay of the waterfall field. Either the waterfall sector decays directly into SM particles or its decay products interact with some of the latter. We note that the waterfall fields may be charged under gauge symmetries, in which case the relevant terms in the superpotential are of the form $\Phi X_\pm\bar{X}_\pm$, etc, where $X_\pm$ and $\bar{X}_\pm$ transform in conjugate representations of the gauge group. This will then open up new avenues for model-building in inflaton dark matter scenarios besides those described in the minimal model, which lie, however, outside the scope of this work.


\subsection{Embedding in an inflationary model}

We have explored in the previous sections the dynamics of the incomplete decay of the inflaton field within a (longer) reheating period controlled by a distinct waterfall sector. The difference with respect to the minimal model is the additional entropy production at the end of reheating which
dilutes the inflaton-dark matter abundance, therefore allowing for larger inflaton masses. However, as in the minimal case, evaporation of the
condensate seems unavoidable, leaving us with similar constraints for the inflaton mass as in the previous scenario, i.e.~masses in the GeV-TeV range. 


As for the minimal model, such a low mass for the inflaton field at the minimum constrains the form of the scalar potential that simultaneously yields a consistent inflationary model and accounts for the observed dark matter abundance. In particular, either inflation can occur at a low energy scale or there must be a large hierarchy between the values of the inflaton mass during inflation and at the minimum of the potential.

In the standard SUSY hybrid models with minimal K\"ahler potential, inflation is driven essentially by the constant vacuum energy $V_0= g^2M^4/8$ while the waterfall fields are stabilized at the origin. We may then use the normalization of the scalar curvature power spectrum:
\be
P_{\cal R}^{1/2} = \frac{H}{2 \pi M_P} \frac{1}{\sqrt{2 \epsilon}} \,,
\ee
and the observational value $P_{\cal R} \simeq 5 \times 10^{-5}$ \cite{planck} to obtain the relation:
\be
m_\phi \gtrsim 2.5 \times 10^{15} |\eta| \mathrm{GeV}~,
\ee
for $\phi\gtrsim M$, where $m_\phi=gM$ is the inflaton mass at the minimum. Since the scalar spectral index $n_s \simeq 1 +2\eta\simeq 0.9603 \pm 0.0073$ at 68\%CL \cite{planck} in these scenarios, we conclude that $m_\phi\lesssim \mathrm{TeV}$ cannot yield an observationally consistent model. Typically we have from the normalization of the spectrum $M \sim 10^{13}-10^{16}$ GeV, and then the WIMPlaton scenario requires small couplings $g \sim 10^{-13}-10^{-10}$, which are responsible for the very flat potential during inflation and the scale invariant spectrum. A non-minimal K\"ahler potential can yield the observed spectral index for lower values of the coupling, although at the expense of a slight increase in $M$ \cite{nonminimal,urRehman:2006hu}, which again makes it difficult to achieve the required WIMPlaton mass values. 

An interesting possibility that can realize inflation at low energy scales is to inflate along the waterfall trajectory, which has been shown to yield a sufficiently red-tilted spectrum for small couplings $g$ within the context of both SUSY and non-SUSY hybrid models \cite{Clesse:2010iz, Kodama:2011vs}. Although it has been recently realized that isocurvature contributions yield a too large enhancement of the amplitude of the spectrum unless $N_e \lesssim 60$, the WIMPlaton scenario typically favors a long reheating period, with e.g.~$N_e \sim 30-40$ being allowed by the data \cite{Clesse:2013jra}.

As for the minimal model, a simple way to realize the WIMPlaton scenario is to inflate in a chaotic rather than in a vacuum-dominated regime \cite{Clesse:2008pf,Clesse:2014fwa,Carpenter:2014saa}. In particular, a quartic self-coupling $\lambda^2\phi^4/4$ can be easily introduced by a superpotential coupling between the inflaton and the the driving field, $\lambda \Phi^2Z/2$ as in Eq.~(\ref{superpotential}). We then have:
\begin{equation}
{\lambda^2\phi^4\over V_0}\sim6\times10^{16} \left({\lambda^2\over 10^{-14}}\right)\left({1\ \mathrm{TeV}\over m_\phi}\right)^{2}\left({M_P\over M}\right)^2\left({\phi\over M_P}\right)^4\,,
\end{equation}
so that the quartic term will easily dominate over the vacuum term for super-planckian values and the typical parameters required by the normalization of the spectrum and the WIMPlaton scenario. We note that the vacuum term will come to dominate the energy at small field values, but for $\phi\gtrsim \phi_c$ we have
\be
\eta \sim \lambda^2 \frac{\phi^2}{M^2}\frac{M_P^2}{m_\phi^2} \gtrsim
\lambda^2 \frac{M_P^2}{m_\phi^2}\gg 1~,  
\ee
such that slow-roll inflation never takes place in the small field regime. Inflation may then occur entirely in a chaotic regime, with a non-minimal coupling to gravity or warm inflation yielding observables within the Planck window as described for the minimal model. 

We note that, in supergravity models, such chaotic inflation scenarios can be obtained by considering a non-minimal K\"ahler potential for the inflaton, while taking the canonical one for the other superfields in the model, in particular the $Z$ field. One possibility is to consider a K\"ahler potential with a shift symmetry \cite{shift}, e.g.~$K(\Phi, Z, \ldots)=(\Phi+ \Phi^\dagger)^2/2+ZZ^\dagger+\ldots$, with inflation taking place along the imaginary component of the scalar inflaton.


\section{Conclusion}
 
In this work we have shown that the decay of the inflaton following the inflationary slow-roll regime can be incomplete, such that successful reheating is achieved while leaving a stable remnant that can account for the observed dark matter in the universe. This is achieved by coupling the inflaton field $\phi$ to a pair of fermions $\psi_\pm$ (and/or similarly scalar fields) while imposing a discrete symmetry that simultaneously changes $\phi\rightarrow -\phi$ and interchanges the two fermions, corresponding to the discrete subgroup $C_2\subset \mathbb{Z}_2\times S_2$. This symmetry forbids all inflaton decay channels except for $\phi\rightarrow \psi_\pm \psi_\pm$ if it is preserved in the vacuum state. The inflaton is thus stable at late times if the bare fermion mass $m_f> m_\phi/2$, where $m_\phi$ denotes the inflaton mass at the origin. However, since the physical fermion mass is field-dependent, inflaton decay may occur during the initial oscillations while the field amplitude is sufficiently large. 

We have then shown that this partial decay is sufficient for a thermalized bath of radiation, which includes the fermions and potentially other degrees of freedom, to become the dominant component. Since the oscillating inflaton condensate does not decay completely, it can account for the observed dark matter in the universe for masses parametrically within the GeV-TeV range in scenarios with a single dynamical scalar field. However, by estimating the scattering rate of zero-momentum inflaton particles off thermalized fermions, we concluded that the oscillating condensate will most likely evaporate in parametric regimes where the reheating temperature is above the threshold required for Big Bang Nucleosynthesis. The stable inflaton particles then reach a thermalized state which eventually decouples from the cosmological radiation bath and freezes out as a standard WIMP. 

In this {\it WIMPlaton} scenario, inflaton masses must also lie in the GeV-TeV range to account for the observed dark matter abundance. While these mass values may {\it a priori} seem too low to yield the correct amplitude for the primordial spectrum of curvature perturbations, we have shown that the inflaton mass can be much larger during the slow-roll period than at the minimum of the potential. This is for example the case of large-field chaotic models, where the inflaton mass during inflation is set by a quartic self-interaction at super-planckian values and a bare quadratic term close to the origin. Such models are consistent with the latest CMB anisotropy measurements by Planck if e.g.~non-minimal couplings to gravity or dissipative effects are included.

In the simplest models with a single dynamical field, the inflaton decay products must interact with the Standard Model (SM) degrees of freedom in order to excite them in the thermal bath. We have explored different possibilities for such interactions, including the case where the inflaton decays into right-handed neutrinos or mili-charged particles in a hidden sector. We have also explored the alternative possibility of hybrid inflation models, where a dynamical waterfall sector, which is also charged under the discrete symmetry, is responsible for reheating the universe. While the incomplete decay of the inflaton is still required to allow the relativistic decay products of the waterfall field to dominate the energy balance, in such scenarios the inflaton decay products need not couple directly to the SM fields, which opens up new phenomenological possibilities. 

Since the entropy produced by the waterfall decay dilutes the inflaton condensate's abundance, it could in principle allow for larger masses accounting for the correct dark matter abundance. However, condensate evaporation will also most likely occur in the viable parametric regimes, such that a WIMPlaton scenario with masses in the GeV-TeV range is again the most probable outcome in hybrid models. A chaotic inflation regime could also provide a consistent inflationary model in this case, although there exist viable scenarios for low-scale inflation such as along the waterfall direction.

A possible alternative that may generically allow for small inflaton masses is the curvaton scenario \cite{curvaton, Mazumdar:2010sa}, where the fluctuations of an additional field that is light during inflation dominate the total curvature perturbations when decaying. In the standard realization of this scenario, the curvaton is a late decaying field, and indeed this would bring some additional dilution of the inflaton abundance due to
entropy production. While this may relax the bound on the inflaton mass in the condensate scenario, it would have practically no influence when the condensate evaporates, and the same limits on the WIMPLaton mass apply. 

Our results show that there exist concrete particle physics models where a single dynamical field can drive inflation in the early universe and account for cold dark matter at late times. This is an appealing feature from the theoretical perspective, since it allows one to address two of the most important problems in modern cosmology within the same simple model. Moreover, we have found that consistent cosmological scenarios typically require the masses of both the inflaton and the particles it interacts directly with to lie in the GeV-TeV range, opening up the possibility of testing both inflation and dark matter at present particle colliders such as the LHC, as well as direct or indirect dark matter searches. Finally, this scenario also singles out particular classes of inflationary models that can be tested with CMB experiments such as Planck, which are now reaching unprecedented levels of precision. We thus hope that our work motivates further exploration of the phenomenological and observational 
consequences of the proposed generic mechanism in its several possible implementations.


\acknowledgments

MBG and RC are supported by the ``Junta de Andaluc\'ia" group FQM101 and project P10-FQM-6552. JGR is supported by the FCT grant SFRH/BPD/85969/2012
and partially by the grant PTDC/FIS/116625/2010, the CIDMA strategic project UID/MAT/04106/2013 and the Marie Curie action
NRHEP-295189-FP7-PEOPLE-2011-IRSES. The authors would like to thank the hospitality of the Higgs Centre for Theoretical Physics during the completion
of this work. RC would like to thank the hospitality of the University of Aveiro and the ``Junta de Andaluc\'ia" for supporting this visit.


\begin{thebibliography}{99}

\bibitem{inflation}
  A.~H.~Guth,
  Phys.\ Rev.\  {\bf D23}, 347 (1981);
  A.~Albrecht, P.~J.~Steinhardt,
  Phys.\ Rev.\ Lett.\  {\bf 48}, 1220 (1982);
  A.~D.~Linde,
  Phys.\ Lett.\  {\bf B108}, 389 (1982).

\bibitem{rotationcurves}
A.~Borriello and P.~Salucci, 
Mon.\ Not.\ Roy.\ Astron.\ Soc.\  {\bf 323}, 285 (2001) 
[arXiv:astro-ph/0001082].  

\bibitem{wmap}
E.~Komatsu {\it et al.}  [WMAP Collaboration],
arXiv:0803.0547 [astro-ph].

\bibitem{Ade:2013ktc}
  P.~A.~R.~Ade {\it et al.}  [Planck Collaboration],
  Astron.\ Astrophys.\  {\bf 571} (2014) A1
  [arXiv:1303.5062 [astro-ph.CO]].

\bibitem{sdss}
 M.~Tegmark {\it et al.} 
[SDSS Collaboration],
Astrophys.\ J.\  {\bf 606}, 702 (2004)
[arXiv:astro-ph/0310725].

\bibitem{weaklensing}
H.~Hoekstra, H.~Yee and M.~Gladders,
New Astron.\ Rev.\  {\bf 46}, 767 (2002)
[arXiv:astro-ph/0205205].
  
\bibitem{Taoso:2007qk}
  M.~Taoso, G.~Bertone and A.~Masiero,
  JCAP {\bf 0803} (2008) 022
  [arXiv:0711.4996 [astro-ph]].

\bibitem{Turner:1983he}
  M.~S.~Turner,
  Phys.\ Rev.\ D {\bf 28} (1983) 1243.

\bibitem{Liddle:2006qz}
  A.~R.~Liddle and L.~A.~Urena-Lopez,
  Phys.\ Rev.\ Lett.\  {\bf 97} (2006) 161301
  [astro-ph/0605205].
 
\bibitem{reheating} 
  A. Albrecht, P.J. Steinhardt, M.S. Turner and F. Wilczek,
  Phys.Rev.Lett. {\bf 48}, 1437 (1982);
  A.D. Dolgov and A.D. Linde, Phys. Lett. {\bf 116B}, 329 (1982);
  L.F. Abbott, E. Fahri and M. Wise, Phys. Lett. {\bf 117B}, 29 (1982);
  L.~Kofman, A.~D.~Linde and A.~A.~Starobinsky,
  Phys.\ Rev.\ Lett.\  {\bf 73} (1994) 3195
  [hep-th/9405187].

\bibitem{Kofman:1997yn}
  L.~Kofman, A.~D.~Linde and A.~A.~Starobinsky,
  Phys.\ Rev.\ D {\bf 56} (1997) 3258
  [hep-ph/9704452].
   
\bibitem{Berera:1995wh} 
  A.~Berera and L.~-Z.~Fang,
  Phys.\ Rev.\ Lett.\  {\bf 74}, 1912 (1995) 
  [astro-ph/9501024];
  A.~Berera,
  Phys.\ Rev.\ Lett.\  {\bf 75}, 3218 (1995)
  [astro-ph/9509049].

\bibitem{Berera:2008ar} 
  A.~Berera, I.~G.~Moss and R.~O.~Ramos,
  Rept.\ Prog.\ Phys.\  {\bf 72}, 026901 (2009)
  [arXiv:0808.1855 [hep-ph]].


\bibitem{thermal}
  V.~H.~Cardenas,
  Phys.\ Rev.\ D {\bf 75} (2007) 083512
  [astro-ph/0701624];
  G.~Panotopoulos,
  Phys.\ Rev.\ D {\bf 75} (2007) 127301
  [arXiv:0706.2237 [hep-ph]];
  K.~Mukaida and K.~Nakayama,
  JCAP {\bf 1408} (2014) 062
  [arXiv:1404.1880 [hep-ph]].
\bibitem{th_infl}
  A.~R.~Liddle, C.~Pahud and L.~A.~Urena-Lopez,
  Phys.\ Rev.\ D {\bf 77} (2008) 121301
  [arXiv:0804.0869 [astro-ph]];
 \bibitem{kinetic}
  N.~Bose and A.~S.~Majumdar,
  Phys.\ Rev.\ D {\bf 80} (2009) 103508
  [arXiv:0907.2330 [astro-ph.CO]];
  J.~De-Santiago and J.~L.~Cervantes-Cota,
  Phys.\ Rev.\ D {\bf 83} (2011) 063502
  [arXiv:1102.1777 [astro-ph.CO]];
\bibitem{regeneration}
  N.~Okada and Q.~Shafi,
  Phys.\ Rev.\ D {\bf 84} (2011) 043533
  [arXiv:1007.1672 [hep-ph]];
  A.~de la Macorra,
  Astropart.\ Phys.\  {\bf 35} (2012) 478
  [arXiv:1201.6302 [astro-ph.CO]].
\bibitem{singlet}
  R.~N.~Lerner and J.~McDonald,
  Phys.\ Rev.\ D {\bf 80} (2009) 123507
  [arXiv:0909.0520 [hep-ph]];
  V.~V.~Khoze,
  JHEP {\bf 1311} (2013) 215
  [arXiv:1308.6338 [hep-ph]].


\bibitem{Steigman:1984ac}
  G.~Steigman and M.~S.~Turner,
  Nucl.\ Phys.\ B {\bf 253} (1985) 375.

\bibitem{wukitung} 
  Wu-Ki Tung,
  ``Group Theory in Physics,''
  World Scientific Publishing Co. Pte. Ltd. (1985)
\bibitem{Kolb:1988aj} 
  E.~W.~Kolb and M.~S.~Turner,
  ``The Early Universe. Reprints,''
  Redwood City, USA: Addison-Wesley (1988) 719 P. (Frontiers in physics, 70)


\bibitem{bbn}
K.~A.~Olive, G.~Steigman and T.~P.~Walker,
Phys.\ Rept.\  {\bf 333}, 389 (2000)
[arXiv:astro-ph/9905320].

\bibitem{planck}
 P.~A.~R.~Ade {\it et al.}  [Planck Collaboration],
  Astron.\ Astrophys.\  {\bf 571} (2014) A22
  [arXiv:1303.5082 [astro-ph.CO]].
\bibitem{Ade:2014xna}
  P.~A.~R.~Ade {\it et al.}  [BICEP2 Collaboration],
  Phys.\ Rev.\ Lett.\  {\bf 112} (2014) 241101
  [arXiv:1403.3985 [astro-ph.CO]].

\bibitem{Bartrum:2013fia}
  S.~Bartrum, M.~Bastero-Gil, A.~Berera, R.~Cerezo, R.~O.~Ramos and J.~G.~Rosa,
  Phys.\ Lett.\ B {\bf 732} (2014) 116
  [arXiv:1307.5868 [hep-ph]].

\bibitem{Salopek:1988qh}
  D.~S.~Salopek, J.~R.~Bond and J.~M.~Bardeen,
  Phys.\ Rev.\  D {\bf 40}, 1753 (1989);
  B.~L.~Spokoiny,
  Phys.\ Lett.\  B {\bf 147}, 39 (1984);
  T.~Futamase and K.~Maeda,
  Phys.\ Rev.\  D {\bf 39}, 399 (1989);
  R.~Fakir and W.~G.~Unruh,
  Phys.\ Rev.\  D {\bf 41}, 1783 (1990);
  D.~I.~Kaiser,
  Phys.\ Rev.\  D {\bf 52}, 4295 (1995)
  [arXiv:astro-ph/9408044];
  E.~Komatsu and T.~Futamase,
  Phys.\ Rev.\  D {\bf 59}, 064029 (1999)
  [arXiv:astro-ph/9901127];
  K.~Nozari and S.~D.~Sadatian,
  Mod.\ Phys.\ Lett.\  A {\bf 23}, 2933 (2008)
  [arXiv:0710.0058 [astro-ph]].


\bibitem{Bezrukov:2008dt}
  F.~L.~Bezrukov,
  arXiv:0810.3165 [hep-ph];
  F.~L.~Bezrukov and M.~Shaposhnikov,
  Phys.\ Lett.\  B {\bf 659}, 703 (2008)
  [arXiv:0710.3755 [hep-th]];
  F.~L.~Bezrukov, A.~Magnin and M.~Shaposhnikov,
  Phys.\ Lett.\  B {\bf 675}, 88 (2009)
  [arXiv:0812.4950 [hep-ph]];
  F.~Bezrukov, D.~Gorbunov and M.~Shaposhnikov,
  JCAP {\bf 0906}, 029 (2009)
  [arXiv:0812.3622 [hep-ph]];
  F.~Bezrukov and M.~Shaposhnikov,
  JHEP {\bf 0907}, 089 (2009)
  [arXiv:0904.1537 [hep-ph]].

\bibitem{Okada:2010jf} 
  N.~Okada, M.~U.~Rehman and Q.~Shafi,
  Phys.\ Rev.\ D {\bf 82}, 043502 (2010)
  [arXiv:1005.5161 [hep-ph]].

\bibitem{Aad:2012tfa} 
  G.~Aad {\it et al.}  [ATLAS Collaboration],
  Phys.\ Lett.\ B {\bf 716}, 1 (2012)
  [arXiv:1207.7214 [hep-ex]];
  S.~Chatrchyan {\it et al.}  [CMS Collaboration],
  Phys.\ Lett.\ B {\bf 716}, 30 (2012)
  [arXiv:1207.7235 [hep-ex]].

\bibitem{Vogel:2013raa} 
  H.~Vogel and J.~Redondo,
  JCAP {\bf 1402}, 029 (2014)
  [arXiv:1311.2600 [hep-ph]].

\bibitem{susyhybrid}
 G. R. Dvali, Q. Shafi and R. K. Schaefer, Phys. Rev. Lett. {\bf 73}
 (1994)  1886. 
\bibitem{hybridinf}
A. D. Linde, Phys. Lett. {\bf B 249} (1990) 18; 
A. D. Linde, Phys. Lett. {\bf B 259} (1991) 38; 
A. D. Linde, Phys. Rev. {\bf D 49} (1994) 748; 

\bibitem{hybridinf2} E. J. Copeland, A. R. Liddle, D. H. Lyth,
E. D. Stewart and D. Wands, Phys. Rev.{\bf D 49} (1994)
6410. 

\bibitem{Dterm}
 P. Binetruy and G. Dvali, Phys. Lett. {\bf B388} (1996) 241
 [hep-ph/9606342];
E. Halyo, Phys. Lett. {\bf B387} (1996) 43 [hep-ph/9606423]

\bibitem{BasteroGil:1997vn}
  M.~Bastero-Gil and S.~F.~King,
  Phys.\ Lett.\ B {\bf 423} (1998) 27
  [hep-ph/9709502]; 
  M.~Bastero-Gil and S.~F.~King,
  Nucl.\ Phys.\ B {\bf 549} (1999) 391
  [hep-ph/9806477].


\bibitem{Antusch:2009ef} 
  S.~Antusch, K.~Dutta and P.~M.~Kostka,
  Phys.\ Lett.\ B {\bf 677}, 221 (2009)
  [arXiv:0902.2934 [hep-ph]];
  AIP Conf.\ Proc.\  {\bf 1200}, 1007 (2010)
  [arXiv:0908.1694 [hep-ph]].



\bibitem{turner}
R.~J.~Scherrer and M.~S.~Turner,
  Phys.\ Rev.\ D {\bf 31} (1985) 681.

\bibitem{nonminimal}
  M.~Bastero-Gil, S.~F.~King and Q.~Shafi,
  Phys.\ Lett.\ B {\bf 651} (2007) 345
  [hep-ph/0604198].

\bibitem{urRehman:2006hu}
  M.~ur Rehman, V.~N.~Senoguz and Q.~Shafi,
  Phys.\ Rev.\ D {\bf 75} (2007) 043522
  [hep-ph/0612023].

\bibitem{Clesse:2010iz}
  S.~Clesse,
  Phys.\ Rev.\ D {\bf 83} (2011) 063518
  [arXiv:1006.4522 [gr-qc]]; 
  S.~Clesse and B.~Garbrecht,
  Phys.\ Rev.\ D {\bf 86} (2012) 023525
  [arXiv:1204.3540 [hep-ph]].

\bibitem{Kodama:2011vs}
  H.~Kodama, K.~Kohri and K.~Nakayama,
  Prog.\ Theor.\ Phys.\  {\bf 126} (2011) 331
  [arXiv:1102.5612 [astro-ph.CO]].

\bibitem{Clesse:2013jra}
  S.~Clesse, B.~Garbrecht and Y.~Zhu,
  Phys.\ Rev.\ D {\bf 89} (2014) 063519
  [arXiv:1304.7042 [astro-ph.CO]].






\bibitem{Clesse:2008pf}
  S.~Clesse and J.~Rocher,
  Phys.\ Rev.\ D {\bf 79} (2009) 103507
  [arXiv:0809.4355 [hep-ph]].


\bibitem{Clesse:2014fwa}
  S.~Clesse and J.~Rekier,
  Phys.\ Rev.\ D {\bf 90} (2014) 8,  083527
  [arXiv:1407.1984 [astro-ph.CO]].

\bibitem{Carpenter:2014saa}
  L.~M.~Carpenter and S.~Raby,
  Phys.\ Lett.\ B {\bf 738} (2014) 109
  [arXiv:1405.6143 [hep-ph]].

\bibitem{shift} 
  M.~Kawasaki, M.~Yamaguchi and T.~Yanagida,
  Phys.\ Rev.\ Lett.\  {\bf 85} (2000) 3572
  [hep-ph/0004243]; 
  P.~Brax and J.~Martin,
  Phys.\ Rev.\ D {\bf 72} (2005) 023518
  [hep-th/0504168]; 
  R.~Kallosh and A.~Linde,
  JCAP {\bf 1011} (2010) 011
  [arXiv:1008.3375 [hep-th]].


\bibitem{curvaton}
D. H. Lyth and D. Wands, Phys. Lett. {\bf B 524} (2002) 5 
[arXiv:hep-ph/0110002];
T. Moroi and T. Takahashi, Phys. lett. {\bf B522} (2001) 215; Erratum
ibid. B539(2002) 303 [arXiv:hep-ph/0110096]; 
K. Enqvist and M. S. Sloth, Nucl. Phys. {\bf B626} (2002) 395
[arXiv:hep-ph/0109214]. 

\bibitem{Mazumdar:2010sa} 
For a review on curvaton and particle physics models see for example:  
  A.~Mazumdar and J.~Rocher,
  Phys.\ Rept.\  {\bf 497} (2011) 85
  [arXiv:1001.0993 [hep-ph]].



\end{thebibliography}
\end{document}